\documentclass[aps,showpacs]{revtex4}
\usepackage{amssymb}
\usepackage{graphicx}
\usepackage{subfigure}
\usepackage{wrapfig}
\usepackage{amsmath}
\usepackage{hyperref}
\usepackage{amsfonts}
\begin{document}

\title{ Approximate joint measurement of qubit observables through an Arthur-Kelly type model}
\author{Rajarshi Pal }
\email{rajarshi@imsc.res.in}
\author{Sibasish Ghosh}
\email{sibasish@imsc.res.in}
\affiliation{The  Institute of Mathematical Sciences, C.I.T Campus,Taramani, Chennai-600110, India.}
        
\begin{abstract}
We consider joint measurement of two and three unsharp qubit observables through an Arthur-Kelly type joint measurement model for qubits. We investigate the effect of initial state of the detectors on the unsharpness of 
the measurement as well as the post-measurement state of the system. Particular emphasis  is given on a physical understanding of  
the POVM to PVM transition in the model and entanglement between system and detectors.Two approaches for characterizing the  unsharpness of the  measurement and the resulting 
measurement uncertainty relations are considered.The corresponding measures of 
unsharpness are connected for the case where both the measurements are equally unsharp. The connection between the POVM elements and symmetries of the underlying Hamiltonian of the measurement interaction is made explicit and used to perform joint measurement
 in arbitrary directions. Finally in the   case of  three observables  we derive a necessary condition for the approximate joint measurement and use it show the relative freedom available when the observables are non-orthogonal.      
\end{abstract}

\maketitle

\section{Introduction}
Quantum mechanics does not allow the joint measurement of non-commuting observables. The uncertainty principle, usually described as lower bound on the product of standard deviations of the outcome statistics does 
not really capture this complementary feature as it (uncertainty principle)  is a statement about the quantum state of the system and does not relate to an actual situation involving apparatus which can attempt a joint measurement.

   As intuition suggests one has to allow for some degree of imprecision in order to make room for a notion of joint measurement of non-commuting observables. In quantum mechanics of a single system observables correspond to self-adjoint operators 
with the outcome statistics given by the Born rule , i.e if $P_i$ is the projector onto the eigenspace corresponding to the eigenvalue $\lambda_i$ then in a measurement of the observable the probability of obtaining the ith outcome when the system is
 in state $\rho$ is $ Tr(\rho P_i) $. However this is not adequate. For example, suppose a system is allowed to interact with another system(ancilla) for a while and then measurement is done on the combined system. To describe the  
 probabilities of obtaining various outcomes for such a measurement for different states of  the original system before the interaction one has to replace the projectors above 
by positive operators $E_i$  (acting on the states of the original system) with $\sum_i E_i =I $ for normalization. Such measurements are called POVM (positive operator valued measures). The most general 
measurements possible in quantum mechanics can be described through POVMs. 
It turns out that the idea of imprecise or unsharp measurements can be appropriately investigated if one considers instead of projective measurements, POVMs. The observables corresponding 
to projective measurements form a subclass of those described by POVMs and are called sharp.

       The first model for approximate joint measurement of position and momentum was given by Arthur and Kelly in (1965) by generalizing von Neumann's model for measurement.(\cite{HBK-2})
In the von Neumann's model for measurement, the position observable of the object is measured by coupling it to the momentum $P_p$ of a probe system via the interaction evolution $ U=e^{-\frac{i}{h}\lambda Q \otimes P_p }$ and using the position $Q_p$ of the probe as the 
readout observable. The idea of Arthur-Kelly was to couple two such probe systems respectively to the position and momentum of the system and then perform measurement on the commuting meter observables of the probe to gain information
 about the  position and momenta of the system. This is in fact an unsharp joint measurement and inspired the later development of the formalism . EPR-Bell argument for unsharp realities was given in \cite{busch-unsharp}; 
the bounds on joint measurement of qubit observables were derived starting from the  condition of operational locality in \cite{operational-locality}; the formalism for qubit observables and relevant inequalities for the
two observable  case were developed in \cite{ajm}, \cite{busch-schmidt},\cite{coexistence}, \cite{oh}. The connection of approximate  joint  measurement with Bell inequalities were investigated in \cite{ajm-bell}; 
 joint measurements by quantum cloning has been  investigated for example in \cite{cloning-barnett} and \cite{cloning-dariano}. It has also been used to estimate the expectation values of qubit observables in \cite{necessity-erica};
 joint unsharp measurement of position and momentum  was treated in  \cite{busch-physics} and \cite{Werner-qp}.                

       In this paper we consider an Arthur-Kelly like model for approximate joint measurement separately, of two and  three non-commuting qubit observables . The probe systems we employ are continuous 
as typically in experiments like Stern-Gerlach deviations in position or momenta of particles carrying spin are  measured to obtain information about the system spin state. We consider characterisation of 
unsharpness or quality of approximation by two ways existing in literature. Firstly by considering closeness of the marginal probability distribution of the probability distribution of joint measurement 
to that of the sharp observable being approximately measured .Secondly by 
considering suitably defined  closeness of the observables themselves with the meter observables in the Heisenberg picture. We show that for a symmetric joint measurement where the two marginal probability distributions 
are equally close to the corresponding  sharp probability  distributions, the two measures of closeness are proportional. Error-disturbance relationship (\cite{appleby-3})  does not seem to hold for  the measure based
on the Heisenberg picture for our choice of the pointer observable. Numerical analysis is performed for the two  observable case to show 
the validity of measurement uncertainty relations, transition  from POVM to projection-valued measurement and also the effect of the joint measurement on the system.
The effect of the pre-measurement on the system turns out to be that of an asymmetric depolarising channel 
and this forms the basis for a physical understanding of the POVM to PVM transition. Entanglement between system and detectors is also investigated .
We also prove a lemma showing the connection between joint measurement ,and the symmetries of the underlying Hamiltonian 
of the measurement interaction together with that of the initial detector states. This is then used to perform approximate joint measurement in arbitrary directions. Moving on to the case of three-observable joint unsharp measurement we  
prove a simple necessary condition that is sufficient for the case of three orthogonal observables. The case of three orthogonal observables also generalises the validity of the known 
necessary-sufficient condition for this case. This necessary condition is derived from certain geometrical considerations based on the so called Fermat-Toricelli point. We show that      
it yields the known two-observable bounds in the limit that the measurement of one of the observables is pure guessing of the value of the observable. 
Finally, an extension of the Arthur Kelly like model to the three observable case is studied. 

      The paper is organised as follows . In section II we introduce the joint unsharp spin measurement and the original Arthur-Kelly model given for the joint measurement of position and momentum.
Measures characterising the quality of approximations to the sharp observables being approximated are introduced in section III . The corresponding measurement uncertainty relations are discussed. 
In section IV we introduce the Arthur-Kelly like model for qubits that we have considered and derive the final state of the system and meters after the measurement interaction. Approximate joint measurement  
of $\sigma_x$ and $\sigma_y$ is considered in section V.   
Section VI deals with the  effect of the initial detector states on the joint measurement and post-measurement state of the system. 
In section VII we try to develop a physical understanding of the POVM to PVM transition seen in the model based on the results obtained in the 
previous section. We also try to see how the entanglement between system and detectors behave as the measurements become sharper. 

In section VIII we explore through a lemma the connection between the symmetries of the 
underlying Hamiltonian of the measurement interaction and initial detector states on the joint measurement. This is used in section IX
 to do  joint measurement of spin in arbitrary directions.The corresponding POVM 
elements are also calculated and matched with the orthogonal case  in section V. In section X we compute the spin direction fidelities, 
which were introduced as a measure of quality of approximation in section III, for  
the model considered. It is compared with another measure based on distance between outcome probabilities of  sharp observables and their
 unsharp approximations. Approximate joint measurement of 
three qubit observables is considered in section XI. A necessary condition on the parameters of the marginal POVM  elements is derived by geometric considerations involving the Fermat-Toricelli point. The 
sufficiency of the condition is explored and the restrictions placed by it  investigated in the context of joint measurement through  an  extension of the model considered in section IV to three detectors.

\section{An introduction to joint unsharp spin measurements and Arthur-Kelly Model}
Suppose we want to measure the spin  of a spin-$\frac{1}{2}$ particle along a direction given by the unit vector $\hat{n}$ .If the system is described by the density matrix $\rho$, then  probabilities for obtaining outcomes $\pm$ are given 
in the POVM formalism, respectively
 by $Tr(\rho A_{+})$ and $Tr(\rho A_{-})$. The POVM elements $A_+$ and $A_-$ are called effects. One says that the observable A is characterized by specification of the map $\omega_i \to A_i$ where $\omega_i$ belongs to the set of outcomes
 with $i=+,-$ (e.g here $\omega_+ = +$ , $\omega_- = -$) .  If $A_+ = |\hat{n},+\rangle \langle \hat{n},+| $ ,i.e,  a projector on to the up state of $ \vec{\sigma}.\hat{n}$ then the measurement is called a sharp measurement of $ \vec{\sigma}.\hat{n}$ and the observable characterised by $A_+ (A_- = I - A_+)$ a sharp observable. Else the measurement and the corresponding observable is called unsharp.     
            One way such a measurement can occur is the following. Suppose some one wants to do a standard (sharp)measurement of $\vec{\sigma}.\hat{n}$ but because of some error in his setup there is a finite probability of registering the outcome - when the system is actually 
in the state $|+ \rangle$ and vice versa. This will be an unsharp measurement with the probabilities  of error given by $Tr(A_{-}|+\rangle \langle +|)$ and $Tr(A_{+}|-\rangle \langle -|)$. Such a situation can arise if for example in a 
Stern-Gerlach setup , the beam passing through the inhomogeneous magnetic field is poorly collimated initially. (\cite{Myunck} and \ref{sec-physics} )  

\subsection{Joint measurability }            
Two observables are said to be jointly measurable if there is a measurement scheme that allows the determination of values of both the observables. This means that the POVM describing the measurement scheme contains 
the POVM elements of the two  observables as marginals. In this way it is ensured that there is a joint probability distribution corresponding to the pairs of different values of the two observables for each state.
                 Two observables $\hat{\Upsilon}_1$ and $\hat{\Upsilon}_2$ are jointly measurable if there is an observable $G:\omega_{ij} \to G_{ij} , i,j=\pm,$ such that,
\begin{eqnarray}          
\label{jointpovm}
\hat{\Upsilon}^1_+ &=&  G_{++} + G_{+-} \nonumber\\
\hat{\Upsilon}^2_+ &=&  G_{++} + G_{-+} \nonumber\\
\hat{\Upsilon}^1_- &=&  G_{-+} + G_{--} \nonumber\\
\label{eqjoint}
\hat{\Upsilon}^1_- &=&  G_{+-} + G_{--}
\end{eqnarray}

with the POVM elements $\hat{\Upsilon}^j_{\pm} (\mbox{for } j=1,2)$ satisfying , $\hat{\Upsilon}^1_+ + \hat{\Upsilon}^1_- = I$ and $\hat{\Upsilon}^2_+ + \hat{\Upsilon}^2_- = I$  . The outcomes $\omega_{ij}$ of G can be taken to be the pairs $(\omega_i , \omega_j) $. The map implies that  
the probability of registering outcome , say  (+,+) is given by $Tr(\rho G_{++} )$ and so on. The extension to three observables is done in the obvious way. 

 Two observables $\hat{\Upsilon}_1$  and $\hat{\Upsilon}_2$ are said to commute if $ \hat{\Upsilon}^1_i \hat{\Upsilon}^2_j=\hat{\Upsilon}^2_j \hat{\Upsilon}^1_i \mbox{ }\forall i,j=\pm  $. 
 The possibility of joint measurability is guaranteed by the following theorem which we state without proof.(\cite{lahti})

\textbf{Theorem 1}: A pair of sharp observables is jointly measurable iff they commute . Commutativity of unsharp observables is  sufficient but not necessary for joint measurability.

\subsection{Arthur-Kelly Model}          
 
As mentioned earlier Arthur-Kelly gave a model for joint measurement
of position and momentum.\cite{HBK-2} In this model a quantum object is coupled with two probe systems which are then individually measured to obtain information about objects position and momentum. 
The coupling to probes is based on von-Neumann's model of measurement. They showed that this constitutes a simultaneous measurement of position and momentum in the sense that the output statistics reproduce the expectation 
values of the object's position and momentum. 

 The position $\hat{Q}$ and momentum $\hat{P}$ of the object are coupled with the position $\hat{Q_1}$ and the momentum $\hat{P_2}$ of the two probe systems respectively which serve as the readout observables.
Neglecting the free evolutions of the three systems (assuming that the measurement interaction dominates during the short time in which it acts) the combined time evolution is described by ,
\begin{equation} 
\label{eqakinteraction}
U=exp^{(-i\lambda \hat{Q} \otimes \hat{P_1} \otimes I_2  + i\kappa \hat{P} \otimes \hat{I_1} \otimes Q_2  )}
\end{equation}
           The coupling constants $\lambda$ and $\kappa$ represent the interaction strength and can be absorbed into a rescaling of the pointer observables $\hat{Q}_1$ and $\hat{P}_2$.
          If $|\psi \rangle$ is an arbitrary input state of the object and $|\Psi_1 \rangle$ and $|\Psi_2 \rangle$ are fixed initial state of the probes (well-behaved and zero expectations for each of the probe's position and momentum ) , the probabilities for values of $\hat{Q}_1$ 
and $\hat{P}_2$ to lie in intervals  $\lambda X$ and $\kappa Y$ are determined through a joint observable  $G^{T}$ \cite{busch-physics} ,analogous to the one defined in eqn. (\ref{eqjoint}) , through
\begin{equation} 
\langle \psi |G^T(X \times Y)|\psi \rangle := (\langle \psi| \otimes \langle \Psi_1 | \otimes \langle \Psi_2|)U^{\dagger}( I \otimes \hat{Q}_1(\lambda X) \otimes \hat{P}_2(\kappa Y) ) U(|\psi \rangle \otimes |\Psi_1 \rangle \otimes | \Psi_2 \rangle) .
\end{equation}
The marginals are given by $G_1^T(X) := G^T(X \times \cal{R})$ and  $G_2^T(Y) := G^T(\cal{R} \times Y)$ .The cost of the joint measurement
 is an increase in the variance of the marginals given by,
\begin{equation}
\Delta(G_1^T,\psi)\Delta(G_2^T,\psi) \geq \hbar
\end{equation}
where, $\Delta(G_j^T,\psi) = \sqrt{\langle \psi |{G_j^T}^2 |\psi \rangle -  {\langle \psi |G_j^T | \psi \rangle}^2}$ for $j=1,2$ .
\section{Quality of joint measurement and measurement uncertainty relations}
The notion of approximate joint measurement naturally demands a measure of proximity to the sharp observables being approximately measured.
 The restriction on the measures corresponding to non-commuting observables leads one to measurement uncertainty relations. In this paper we use two such approaches to characterise  proximity.
\subsection{Closeness of probabilities}
Proximity between two observables can be characterised by the distance between the corresponding probability distributions for all states. 
Thus distance between two observables $\hat{A}$ and $\hat{B}$ can be defined as   
\begin{equation} 
D (\hat{A},\hat{B}) := {\mbox{max}}_{j} {\mbox{sup}}_{T} |tr[TA_j] - tr[TB_j]|  
\end{equation}
where $A_j(or B_j)$ corresponds to the POVM element(or effect) of the observable A (or B) associated with the measurement outcome $j $ . $T$ is the density matrix of the system.
\cite{ajm} 

For two single-qubit observables $\hat{\Upsilon}^{\alpha,\vec{a_1}}$ ,$\hat{\Upsilon}^{\beta,\vec{a_2}}$ with,
\begin{eqnarray}
\hat{\Upsilon}^{\alpha,\vec{a_1}} = \frac{1}{2}(\alpha I + \vec{a_1}.\vec{\sigma}) \\
\hat{\Upsilon}^{\beta,\vec{a_2}} = \frac{1}{2}(\beta I + \vec{a_2}.\vec{\sigma})
\end{eqnarray}
with $(\alpha,\vec{a_1})$ and $(\beta,\vec{a_2}) \in {\cal{R}}^4$ . Now for observables $\hat{A}$ , $\hat{B}$ with respective 
set of effects  $\{A_+ , A_-\}$ , $\{B_+ , B_-\}$ we have $|tr[TA_+] - tr[TB_+]|= |tr[TA_-] - tr[TB_-]|$. Taking T to be the $'+'$  state or $'-'$ state of 
$\vec{\sigma}.\frac{(\vec{a_1}-\vec{a_2})}{||(\vec{a-1}-\vec{a_2})||}$  according as whether $(\alpha-\beta) > 0 $  or  $(\alpha-\beta) < 0$ respectively one has,
\begin{equation}
D(\hat{\Upsilon}^{\alpha,\vec{a_1}},\hat{\Upsilon}^{\beta,\vec{a_2}})= \frac{1}{2} ||\vec{a_1}-\vec{a_2}|| + \frac{1}{2}|\alpha - \beta|
\end{equation}
This shows that the distance of a certain unsharp observable $\hat{\Upsilon}^{\alpha,\vec{a_1}}$ from any sharp observable 
$\hat{\Upsilon}^{1,\hat{n}}$ is minimum when $\vec{a_1}$ is along $\hat{n}$.  

\subsubsection{Unbiased Observables}
\label{subsection-unbiased}
Observables  of the form  $\hat{\Upsilon}^{1,\vec{a}}$ are called unbiased.
 As $\hat{\Upsilon}^{1,\vec{a}} = \frac{(1+|\vec{a}|)}{2}\frac{1}{2}(I+ \frac{\vec{a}.\vec{\sigma}}{|\vec{a}|}) + \frac{(1-|\vec{a}|)}{2}\frac{1}{2}(I- \frac{\vec{a}.\vec{\sigma}}{|\vec{a}|})$, therefore the probability of occurence of outcomes
+(or -) for the initial state $\frac{I}{2}$ (the maximally mixed state) is given by, $Tr [\frac{I}{2}.\frac{1}{2}(I+ \frac{\vec{a}.\vec{\sigma})}{|\vec{a}|}] = \frac{1}{2} 
 \left( \mbox{or  } Tr [\frac{I}{2}.\frac{1}{2}(I- \frac{\vec{a}.\vec{\sigma})}{|\vec{a}|})] = \frac{1}{2} \right )  $. So these two probabilities are  same. Hence the name 'unbiased'.

 For such an  observable, both the outcomes are equally likely for a maximally mixed state . Also, the expectation value of the unsharp measurement when the system
is in state T is given by ${\langle \vec{\sigma}.\hat{a} \rangle}_{u} := 1.tr(T \frac{1}{2}(I + \vec{a}.\vec{\sigma})) -1.tr(T \frac{1}{2}(I - \vec{a}.\vec{\sigma})) = ||\vec{a}||\langle \vec{\sigma}.\hat{a} \rangle $. Again as, 
$D(\hat{\Upsilon}^{1,\hat{a}},\hat{\Upsilon}^{1,\vec{a}}) = \frac{1}{2}(1-||\vec{a}||)$, $||\vec{a}||$ itself serves as a measure of proximity . We will often approximate sharp observables with unbiased unsharp ones.

\subsubsection{Measurement uncertainties}
We will choose jointly measurable observable pairs $(\hat{\Upsilon}^{\alpha,\vec{a}},\hat{\Upsilon}^{\beta,\vec{b}})$ to approximate the sharp pair $(\hat{\Upsilon}^{1,\hat{n}},\hat{\Upsilon}^{1,\hat{m}})$.
 The necessary and sufficient conditions on $\alpha ,\beta ,\vec{a},\vec{b}$ so that the first pair is jointly measurable in the sense of \ref{eqjoint} is formulated in \cite{coexistence},\cite{busch-schmidt}, \cite{oh}.
        When the observables are unbiased the conditions simplify to,
\begin{equation}
\label{eqajmuncer}
||\vec{a}+ \vec{b}|| + ||\vec{a}- \vec{b}|| \leq 2 
\end{equation}
        It was further shown in \cite{ajm} that,
\begin{equation}
\label{mur}
D(\hat{\Upsilon}^{\alpha,\vec{a}},\hat{\Upsilon}^{1,\hat{n}}) + D(\hat{\Upsilon}^{\beta,\vec{b}},\hat{\Upsilon}^{1,\hat{m}}) \geq 2D_0
\end{equation}
with, $D_0= \frac{1}{\sqrt{2}}(cos(\theta/2)+ sin(\theta/2)-1)$ , $\theta$ being the angle between $\hat{a}$ and $\hat{b}$ . The conditions for attainment  of the 
lower bound $2D_0$ were also spelt out . The approximate observables should be unbiased, i.e 
$\alpha=\beta=1$ in this later case. The other conditions(\cite{ajm}) imply that for optimality,  $\vec{a}$ and $\vec{b}$  lie along $\hat{n}$ and $\hat{m}$ respectively, only when the latter are orthogonal. 

\subsection{Closeness of observables}
A completely different approach was taken in references \cite{HBK-1} and \cite{appleby-2} in the context of the original Arthur-Kelly model to give a formulation of the complementary nature of the approximate joint measurement process.
       If $\hat{\mu}_x$ and $\hat{\mu}_p$ denote the pointer observables used to measure the system position and momentum respectively , then the  retrodictive error operators are defined as ,
\begin{eqnarray}
\hat{{\epsilon_X}_i} = \hat{{\mu_x}}_f - \hat{x_i}, 
\mbox{    }\hat{{\epsilon_P}_i}= \hat{{\mu_p}}_f - \hat{p_i}
\end{eqnarray}
the predictive error operators as,
\begin{eqnarray}
\hat{{\epsilon_X}_f} = \hat{{\mu_x}}_f - \hat{x_f}, 
\mbox{    }\hat{{\epsilon_P}_f}= \hat{{\mu_p}}_f - \hat{p_f}
\end{eqnarray}
and the disturbance operators as,
\begin{eqnarray}
\hat{\delta}_X = \hat{x_f} - \hat{x_i}, 
\mbox{    }\hat{\epsilon}_P= \hat{p_f} - \hat{p_i}
\end{eqnarray}
where operators $\hat{O_f}$, appearing on right stands for the final Heisenberg picture operator $\hat{O}$ after the measurement interaction U, i.e, $\hat{O_f}= U^{\dagger}O_iU$ where $\hat{O}_i$ is the Heisenberg picture operator at the moment the interaction starts.
        Various errors were then defined by taking the square root of expectation of the square of the operators defined above and taking supremum over the system states. For example the maximal error of retrodiction was defined as ,
\begin{equation}
{\Delta_{ei}}_x = {sup}_{|\psi\rangle \in H_{sys}} (\langle \psi \otimes \Psi_1 \otimes \Psi_2 | {\hat{{\epsilon_X}_i}}^2 |\psi \otimes \Psi_1 \otimes \Psi_2 \rangle )^\frac{1}{2},
\end{equation}
 and similarly for  ${\Delta_{ei}}_p$ , ${\Delta_{ef}}_x$, ${\Delta_{ef}}_p$ , ${\Delta_{d}}_x$ and  ${\Delta_{d}}_p$(see \cite{HBK-1} , \cite{appleby-2} for details) .Measurement uncertainty or 'error principle' was shown to hold in the form of,
\begin{eqnarray}
 {\Delta_{ei}}_x {\Delta_{ei}}_p   \geq \frac{\hbar}{2}, \\
 {\Delta_{ef}}_x {\Delta_{ef}}_p   \geq \frac{\hbar}{2} ,\ 
 {\Delta_{ei}}_x {\Delta_{d}}_p   \geq \frac{\hbar}{2} , \\
 {\Delta_{ef}}_x {\Delta_{d}}_p   \geq \frac{\hbar}{2} ,
\end{eqnarray}
and extensions of the above uncertainties in the obvious way.
       One of the important features of this approach is the difference between error of retrodiction and that of prediction .It was shown that these are not the same as long as there is a finite disturbance. 
\subsection{Qubit Observables}
In a similar spirit, for the case of approximate joint measurement of qubit observables through an Arthur-Kelly like process 
fidelities were defined in the Heisenberg picture that would provide a notion of direction of spin \cite{appleby-3} of the system. 
As in the previous case, distinction was made between errors of retrodiction and prediction. 
      In this paper we consider only the type 1 measurements considered by the author of \cite{appleby-3} .  
           We next consider the fidelities as defined by the authors. The retrodictive fidelity is defined as,
\begin{equation}  
\eta_i= {inf}_{|\chi \rangle \in H_{sys}} (\langle \psi| \otimes \langle \chi | \frac{1}{2} ( \hat{n_f}. \hat{S_i} + \hat{S_i}. \hat{n_f})| \psi \rangle \otimes |\chi \rangle )
\end{equation}
where, ${\hat{S_i}= \hat{S}}$ and ${\hat{n_i}= \hat{n}}$ are the initial values of the Heisenberg spin and pointer observables respectively and 
$\hat{n_f}= U^\dagger (\hat{n_i} \otimes 1_s) U$ and $\hat{S_f}= U^\dagger (I \otimes \hat{S_i}) U$ respectively be the final 
Heisenberg pointer and spin direction observables after the measurement interaction.

The predictive fidelity is defined as,
\begin{equation}  
\eta_f= {inf}_{|\chi \rangle \in H_{sys}} (\langle \psi| \otimes \langle \chi | \frac{1}{2} ( \hat{n_f}. \hat{S_f} + \hat{S_f}. \hat{n_f})| \psi \rangle \otimes |\chi \rangle ).
\end{equation}

The measurement disturbance by,
\begin{equation}  
\eta_d= {inf}_{|\chi \rangle \in H_{sys}} (\langle \psi |\otimes \langle \chi | \frac{1}{2} ( \hat{S_f}. \hat{S_i} + \hat{S_i}. \hat{S_f})| \psi \rangle \otimes |\chi \rangle ).
\end{equation}
     The intuition behind the above definitions is classical in the sense that it considers alignment of initial or final spin vector and  initial or final pointer direction.
But  the above fidelities were used to define  maximal rms error of retrodiction,
\begin{eqnarray}
\label{retroerror}
\Delta_{ei}S = {sup}_{|\chi \rangle \in H_{sys}}{({\langle \psi |\otimes \langle \chi||\eta_i \hat{n_f}-\hat{S_i}|}^2 | \psi \rangle \otimes |\chi \rangle )}^\frac{1}{2} \nonumber \\ 
= {(s+s^2- {\eta}_i^2)}^{\frac{1}{2}},
\end{eqnarray}
maximal rms error of prediction,
\begin{eqnarray}
\Delta_{ef}S = {sup}_{|\chi \rangle \in H_{sys}}{(\langle \psi |\otimes \langle \chi|{|\eta_f \hat{n_f}-\hat{S_f}|}^2 |   \psi \rangle \otimes |\chi \rangle )}^\frac{1}{2} \nonumber \\             
= {(s+s^2- {\eta}_f^2)}^{\frac{1}{2}},
\end{eqnarray}
and maximal rms disturbance 
\begin{eqnarray}
\Delta_{d}S = {sup}_{|\chi \rangle \in H_{sys}}{(\langle \psi |\otimes \langle \chi|{| \hat{S_f}-\hat{S_i}|}^2 |  \psi \rangle \otimes |\chi \rangle )}^\frac{1}{2} \nonumber\\ 
= \sqrt{2}{(s+s^2- {\eta}_d^2)}^{\frac{1}{2}},
\end{eqnarray}
where the spin $s=\frac{1}{2}$ for our case.
     The quantities $\Delta_{ei}S$ , $\Delta_{ef}S$, $\Delta_{d}S$ were expected to play the same role for these measurements as similar quantities defined for the original Arthur-Kelly model.  
     The authors in ref.\cite{appleby-3} also showed that the retrodictive and predictive fidelities reach their optimal values of $s=0.5$ when the Krauss operators of the measurement POVM have a spin-coherent form defined using spin coherent states. A physical model for such a measurement was given in  \cite{dariano-1}.

       No measurement uncertainties were derived in \cite{appleby-3}  and the question was left open for further investigation.

\section{The Model}
      We consider an instantaneous coupling interaction with the help of the Hamiltonian of the form, 
\begin{equation}      
\label{model}
H=  -(\hat{q}_1 \otimes \sigma_x +  \hat{q}_2 \otimes \sigma_y)\delta t 
\end{equation}
(in $\hbar$=1 units).
 Possible coupling constants in the above equation like $\lambda$ and $\kappa$ in equation \ref{eqakinteraction} have been absorbed by rescaling $q_1$ and $q_2$.  
Like the original Arthur-Kelly interaction (\ref{eqakinteraction}) the idea is to entangle the detectors with the system through $H$ and then perform a projective measurement of $\hat{p}_1$ , $\hat{p}_2$ to obtain the spin information.
Now, as a consequence of the Ehrenfest theorem the average momentum change of a particle carrying spin that experiences  the above interaction is given by $<\dot{\hat{p}}_1> = <\sigma_x>$ and   $<\dot{\hat{p}}_2> = <\sigma_y>$ . 
Thus, for an ensemble of particles whose spin state is $|+x>$ and which has a symmetric distribution of $p_1$ before the interaction will have a greater probability of  
having a positive $p_1$ after it.The signs in  equation (\ref{model})  
have been chosen so that this fact is true for both x and y directions. The signs thus  allow us to map the four quadrants of the momentum plane ($p_1,p_2$) to the four outcomes of joint measurement and take the signs of the momenta 
to correspond to the outcomes (+,+) , (+,-),(-,+) ,(-,-) of the joint measurement. Note that for a Stern-Gerlach situation the two terms in eqn. (\ref{model}) should have opposite signs to satisfy divergenceless of magnetic field.   

       Models similar to above have been considered before for example in \cite{Myunck},\cite{kienzler},\cite{reality-helicity}. As shown in \cite{kienzler} this model naturally arises in the context of a Stern-Gerlach experiment 
with a linear magnetic field. 

         We further assume that the measurement interaction (\ref{model}) is strong enough to dominate the other parts of the Hamiltonian during its presence (e.g, the kinetic energy part ).  
In the Stern-Gerlach context this would mean to assume the atoms carrying spin to be sufficiently massive. 

 The unitary evolution corresponding to the Hamiltonian of equation (\ref{model}) is given after integrating the time evolution operator by,

\begin{equation}
\label{unitary}
U=exp(i( \hat{q}_1 \otimes \sigma_x + \hat{q}_2 \otimes \sigma_y))
\end{equation}

On direct expansion,  U turns out to have the simple form,
\begin{eqnarray}
U & = & e(\hat{q}_1 , \hat{q}_2 ) \otimes 1_s + f(\hat{q}_1 , \hat{q}_2) \otimes \sigma_x + g(\hat{q}_1 , \hat{q}_2) \otimes \sigma_y \nonumber\\
\mbox{ with,  } e(\hat{q}_1,\hat{q}_2) & = & cos (\sqrt{(\hat{q}_1 ^ 2 + \hat{q}_2 ^2 )}) \nonumber\\
f(\hat{q}_1 , \hat{q}_2) & = & i\hat{q}_1 \frac{sin (\sqrt{(\hat{q}_1 ^ 2 + \hat{q}_2 ^2 )})}{\sqrt{(\hat{q}_1 ^ 2 + \hat{q}_2 ^2 )}}\nonumber \\
g(q_1 , q_2) & =  & i\hat{q}_2 \frac{sin (\sqrt{(\hat{q}_1 ^ 2 + \hat{q}_2 ^2 )})}{\sqrt{(\hat{q}_1 ^ 2 + \hat{q}_2 ^2 )}} 
\end{eqnarray} 

Thus, the final state of the system is given by
\begin{eqnarray}
|{\psi}_f \rangle  = & \int_{q_1,q_2 = -\infty}^{+\infty} e(q_1,q_2) |q_1,q_2 \rangle {\psi}_1(q_1) {\psi}_2(q_2) dq_1 dq_2 \otimes |{\chi}\rangle \nonumber\\ + & \int_{q_1,q_2 = -\infty}^{+\infty} f(q_1,q_2) |q_1,q_2 \rangle {\psi}_1(q_1) {\psi}_2(q_2) dq_1 dq_2 \otimes \sigma_x|{\chi}\rangle \nonumber\\+ & \int_{q_1,q_2 = -\infty}^{+\infty} g(q_1,q_2) |q_1,q_2 \rangle {\psi}_1(q_1) {\psi}_2(q_2) dq_1 dq_2 \otimes \sigma_y|{\chi}\rangle,
\end{eqnarray}
with initial state being, $ |{\psi}_i \rangle = |{\psi}_1 \rangle \otimes |{\psi}_2 \rangle \otimes |{\chi}\rangle $. 
Let, $\rho = | \psi_f \rangle \langle \psi_f |$ .

\section{ Approximate Joint measurement in orthogonal directions}
\label{section-ajmortho}
In this section we consider the joint measurement of $\sigma_x$ and $\sigma_y$.We choose the observables $ \hat{p}_1$ and $\hat{p}_2$ to serve as meters. As mentioned before, 
$(p_1 \geq 0 , p_2 \geq 0)$ is taken to correspond to the outcome $(\sigma_x=1 ,\sigma_y=1) \equiv (+,+)$ of joint measurement, $(p_1 \geq 0 , p_2 \leq 0)$ to  $(\sigma_x=1 ,\sigma_y=-1) \equiv (+,-)$,
 $(p_1 \leq 0 , p_2 \geq 0)$ to  $(\sigma_x=-1 ,\sigma_y=1) \equiv (-,+)$ and
$(p_1 \leq 0 , p_2 \leq 0)$ to $(\sigma_x=-1 ,\sigma_y=-1) \equiv (-,-)$.
              
              After the interaction U between the system and meters , projective measurement is performed separately on the observables $\hat{p}_1$ and $\hat{p}_2$ . The probability of obtaining outcome $(p_1,p_2)$ is
given by,

\begin{align}
p(p_1,p_2) & = Tr(|p_1,p_2 \rangle \langle p_1,p_2 |Tr_s(\rho)) \nonumber\\
& = (|e^0|^2 + |f^0|^2 + |g^0|^2)\langle \chi | \chi \rangle + 2Re(f^0{e^0}^*)\langle \chi | \sigma_x | \chi \rangle  + \nonumber\\
\label{probability}
& 2Re(g^0{e^0}^*)\langle \chi | \sigma_y | \chi \rangle -2Im(g^0{f^0}^*)\langle \chi | \sigma_z | \chi \rangle 
\end{align}
with $e^0$, $f^0$ and $g^0$ representing respectively the fourier transforms of $e\psi_1 \psi_2$, $f\psi_1 \psi_2$ and $g\psi_1 \psi_2$.

The initial pointer states are taken to be Gaussian, 
\begin{eqnarray}
\label{inistate1}
\psi_1(q_1)&=& (\frac{1}{\sqrt{2\pi}}e^-{\frac{q_1^2}{2a^2}})^\frac{1}{2} ,\\  
\label{inistate2}
\psi_2(q_2) &=& (\frac{1}{\sqrt{2\pi}}e^-{\frac{q_2^2}{2b^2}})^\frac{1}{2} .
\end{eqnarray}
satisfying $\int_{-\infty}^{+\infty}|{\psi_j(q_j)}^2| dq_j = 1 $ for $j=1,2$ .

We have chosen the initial states to be even in $q_1$ , $q_2$. Now as  Fourier transform of a real even function is a  real even function and that of an  imaginary odd function is a 
 real odd function we have $f^0$ odd in $p_1$ and even in $p_2 $, $g^0 $the other way around and $e^0$ even in both. Also, each of them is real. Thus, the $\sigma_z$ term in (\ref{probability}) vanishes. Henceforth in this paper we 
refer to these properties of $e^0$, $f^0$ and $g^0$ as ``parity properties''. 

We have for the probability of  outcome (+,+) from equation (\ref{probability}),
\begin{eqnarray}
\label{probp1p2geq0}
p (p_1 \geq 0 , p_2 \geq 0)=& \int_{p_1 =0}^{\infty} \int_{p_2 =0}^{\infty} \{ (|e^0|^2 + |f^0|^2 + |g^0|^2)\langle \chi | \chi \rangle + 2Re(f^0{e^0}^*)\langle \chi | \sigma_x | \chi \rangle   \nonumber \\ 
 & +  2Re(g^0{e^0}^*)\langle \chi | \sigma_y | \chi \rangle  \,  \} dp_1dp_2 .
\end{eqnarray}

One also of course has to satisfy,
\begin{equation*}
p (p_1 \geq 0) + p (p_1 \leq 0) = 1,
\end{equation*}
which yields, due to the ``parity properties'',
\begin{equation}
\label{probnorm}
\int_{p_1 =0}^{\infty} \int_{p_2 =0}^{\infty}  (|e^0|^2 + |f^0|^2 + |g^0|^2) \, dp_1 \, dp_2 = \frac{1}{4} .
\end{equation}

From equation (\ref{probnorm}) and ``parity properties'' we have,
\begin{equation}
\label{jm-povm1} 
p (p_1 \geq 0 , p_2 \geq 0) = \langle \chi |\frac{I}{4}+\frac{a'\sigma_x}{4}+\frac{b'\sigma_y}{4} |\chi \rangle = \langle \chi |G_{++} |\chi \rangle
\end{equation}
with, $G_{++} = (\frac{I}{4}+\frac{a'\sigma_x}{4}+\frac{b'\sigma_y}{4})$ and
\begin{align}
\label{povm-constants}
a' & = \int_{p_1=0}^{+\infty} \int_{p_2=-\infty}^{+\infty} 4(f^0{e^0}) dp_1dp_2 , \\
b' & = \int_{p_2=0}^{+\infty} \int_{p_1=-\infty}^{+\infty} 4(g^0{e^0}) dp_1dp_2  .                      
\end{align}
For the other outcomes, we have from consideration of the corresponding momentum probabilities ,
\begin{eqnarray}
\label{jm-povm2}
G_{+-} &=&  \frac{I}{4}+\frac{a'\sigma_x}{4}-\frac{b'\sigma_y}{4} ,\nonumber \\
G_{--} &=& \frac{I}{4}-\frac{a'\sigma_x}{4}-\frac{b'\sigma_y}{4} ,\nonumber \\
G_{-+} &=& \frac{I}{4}-\frac{a'\sigma_x}{4}+\frac{b'\sigma_y}{4} .
\end{eqnarray}

From equations (\ref{jm-povm1}) and (\ref{jm-povm2}) we get the marginal unsharp observables as, ( see eqn. (\ref{eqjoint}))
\begin{eqnarray}
\label{povm-marginal}
\Upsilon^1_{\pm} &=& \frac{1}{2}(I \pm a'\sigma_x) \nonumber \\ 
\Upsilon^2_{\pm} &=& \frac{1}{2}(I \pm b'\sigma_y)
\end{eqnarray}
Thus, we see that the  approximate observables $\Upsilon^1$ ,  $\Upsilon^2$ are unbiased and  $a'$ and $b'$ themselves serve as measures of proximity to the sharp observables $\frac{1}{2}(I \pm \sigma_x)$ and $\frac{1}{2}(I \pm \sigma_y)$ 
, respectively.  

In section \ref{section-povmsym} we will see how the approximate joint measurement, characterised by the marginals in eqn. (\ref{povm-marginal}) arise as a consequence of symmetries of the Hamiltonian in eqn,(\ref{model}) and that
of the initial detector states $\psi_1$ and $\psi_2$ ( given in eqns. (\ref{inistate1}) and (\ref{inistate2}) ).

\subsection{Rdm of the  system after pre-measurement}
\label{subsection-rdm}
The interaction U (\ref{unitary}) acting on the system and the detectors induce a completely-positive  map on the system . The rdm of the system after the interaction is given by,

\begin{eqnarray}
\rho_f^s &=&  Tr_{1,2} (|\psi_f \rangle \langle \psi_f |) \\  
 &=& \int_{q_1,q_2 = -\infty}^{+\infty} |e|^2 {\psi_1}^2 {\psi_2}^2 dq_1dq_2 |\chi \rangle \langle \chi | + \int_{q_1,q_2 = -\infty}^{+\infty} |f|^2 {\psi_1}^2 {\psi_2}^2 dq_1dq_2 \sigma_x|\chi \rangle \langle \chi |\sigma_x \\
 &+&  \int_{q_1,q_2 = -\infty}^{+\infty}|g|^2 {\psi_1}^2 {\psi_2}^2 dq_1dq_2 \sigma_y|\chi \rangle \langle \chi |\sigma_y \\ 
 &=&  \sum_{j=1}^3 K_j |\chi \rangle \langle \chi| {K_j}^{\dagger}                                                           
\end{eqnarray}

The action of the measurement interaction on the system is thus that of an asymmetric depolarising  channel , with the Krauss operators 
given by $K_1= 2\sqrt{(c_{f})}\sigma_1$ ,   $K_2= 2\sqrt{(c_{g})} \sigma_2$ and $K_3=\sqrt{(1-4c_{f} - 4c_{g})}I $ , where
\begin{eqnarray}
\label{coeff}
c_{f} &=& \int_{q_1,q_2 = -\infty}^{+\infty} |f|^2 {\psi_1}^2 {\psi_2}^2 dq_1dq_2 , \nonumber \\
c_{g} &=&  \int_{q_1,q_2 = -\infty}^{+\infty} |g|^2 {\psi_1}^2 {\psi_2}^2 dq_1dq_2  .
\end{eqnarray}

If $(x,y,z)$ denotes the Bloch vector of $|\chi \rangle \langle \chi |$, then in the Bloch sphere representation we have ,

\begin{equation}
\label{rhofs}
\rho_f^s = \frac{1}{2} (I + x(1-8c_{g})\sigma_x  +  y(1-8c_{f})\sigma_y )
\end{equation}

\section{Effect of initial detector states}

The form of the POVM elements in eqns.(\ref{jm-povm1}) , (\ref{jm-povm2}) which leads to unbiased marginal observables (see section \ref{subsection-unbiased}) only depends on the ``parity properties`` and not on the Gaussian form of the initial detector states given by eqns. (\ref{inistate1}) , 
(\ref{inistate2}).

The integrals for computing  the coefficients $c_{f} $ and $c_{g} $ have been done using the Monte Carlo integrator 
included in GNU Scientific Library. The MISER algorithm has been used which uses stratified random sampling 
for doing importance sampling \cite{miser}.

\section*{Results}
As we are considering unbiased approximate joint measurement in orthogonal directions , eqn. (\ref{eqajmuncer}) applied on the parameters $a'$ and $b'$ of the marginals in eqn. (\ref{povm-marginal}) yield ,
\begin{equation}
\label{ajm-ortho}
{a'^2+b'^2} \leq 1 . 
\end{equation} 

\subsection{Symmetric case}
In this case the standard deviations $a$ and $b$ of the initial momentum wavefunctions in eqns. (\ref{inistate1}) and (\ref{inistate2}) are taken to be equal. As there is nothing else 
that differentiates between the detectors one must have here $a'=b'$ . This is reflected in eqn. (\ref{povm-constants}). Both the observables $\sigma_x$  and $\sigma_y$ are approximated equally well. From eqn.
(\ref{ajm-ortho}) we have, $a' \leq \frac{1}{\sqrt{2}} = 0.707$ .

\subsection*{Quality:}
\begin{figure}
\begin{center}
\subfigure[A plot of $a'$  vrs. $a$]{\label{fig-a'}\includegraphics[height=70mm]{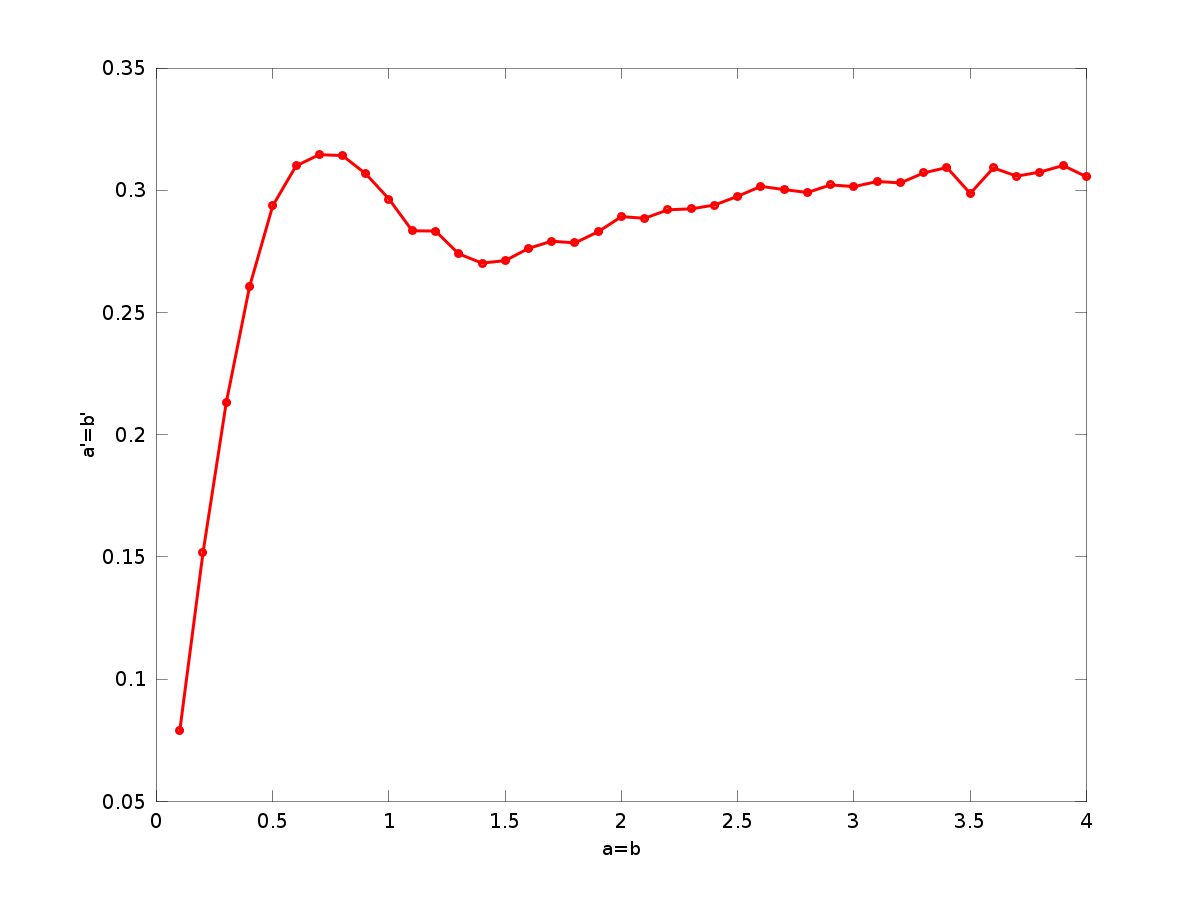}}
\subfigure[A plot of  $1-8c_{f}$ vrs. a ]{\label{fig-distursym}\includegraphics[height=70mm]{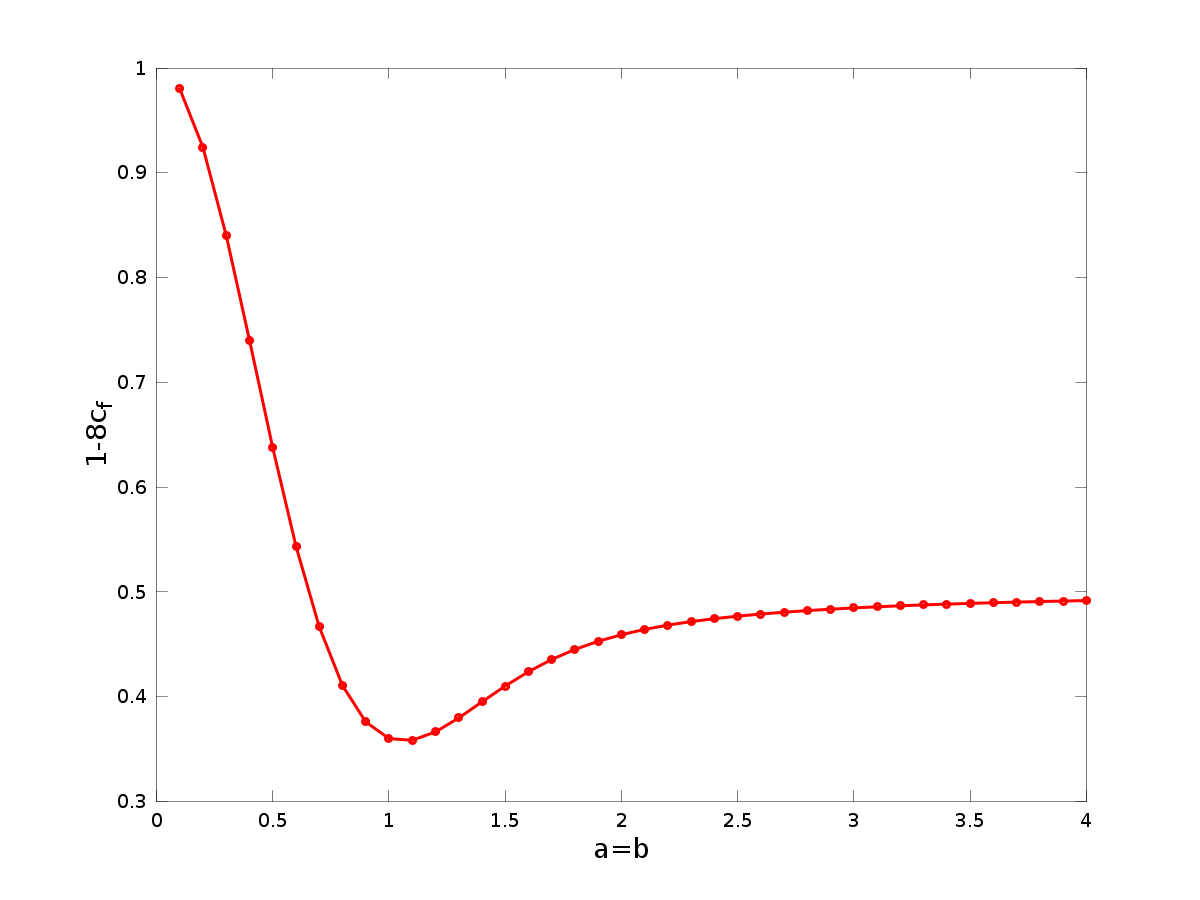}}
\end{center}
\caption{Effects of initial pointer states on the joint measurement and on the system state after the measurement interaction for $a=b$. }
\label{figsym}
\end{figure}

In figure \ref{figsym} we see that when $a$ and $b$ are close to zero, which corresponds to the fact that the 
initial momenta $p_1$ , $p_2$ have large spread we have almost no information about the spin state  by 
looking at the momentum.In this case  $a'$ and $b'$ are both close to zero and we have trivial marginal
 effects ($\frac{I}{2} $) which represent guessing the value of the corresponding sharp observable 
with equal probability for + and -. With increase in $a$, $a'$ increases 
and touches the value 0.628 at  $a=b=0.7$.(see figure \ref{fig-a'}). The graph of $a'$ vrs. $a$ is identical to the one obtained in \cite{kienzler} for the same case. The 
maxima in the curve is a feature of the interaction used and is explained in the next section.

\subsection*{Disturbance due to the measurement interaction}
Figure \ref{fig-distursym} shows the variation of 
 $\frac{{\langle \sigma_x \rangle}_f}{{\langle \sigma_x \rangle}_i} =\frac{{\langle \sigma_y \rangle}_f}{{\langle \sigma_y \rangle}_i}$  
with a,  where ${\langle \sigma_x \rangle}_f = Tr(\rho_s^f\sigma_x)$ and ${\langle \sigma_x \rangle}_i = \langle \chi | \sigma_x | \chi \rangle $ and similarly for $\sigma_y$,
 ( see eqn.(\ref{rhofs})).

 Corresponding to the maximum in $a'$ we also have a  minimum in the disturbance of the 
state characterised by  $\frac{{\langle \sigma_x \rangle}_f}{{\langle \sigma_x \rangle}_i}$ =$\frac{{\langle \sigma_x \rangle}_f}{{\langle \sigma_x \rangle}_i} $ . Thus a sharper measurement seems to disturb 
the state lesser , though there is a slight difference in the value of a at which the maximum and the minimum occur. This disturbance 
is even more prominent in the asymmetric case to be discussed next. 

\subsection*{$b \geq a : \textbf{POVM to PVM transition} $}
a) Let us choose $a=0.1$ . The lhs of eqn.  (\ref{ajm-ortho}) starts off at a low value for $b=a$ and  gradually increases and closes on the bound (1.0) as $b$ becomes much greater than $a$.(see figure \ref{fig-aprbprap1}) 
$ b$ much greater than $a$ reflects the situation where the initial momentum wavefunction of apparatus 2 is much sharper than that of apparatus 1 . As shown clearly by figure \ref{fig-aprbprap1} this marks 
a transition from a POVM measurement to a projection-valued measurement(PVM) in the sense that the unsharp measurement of $\sigma_y$ becomes almost sharp.
The requirement of complementarity is satisfied by the fact that the unsharp  $\sigma_x$ measurement becomes almost trivial. The fact that $a'$ and $b'$ depend both on $a$ and $b$
 is a reflection of the correlations between $p_1$ and $p_2$ brought about by the unitary evolution given by  eqn. (\ref{unitary}).
 
\subsection{Disturbance due to the measurement }          
\label{subsection-distur}
\begin{figure}
\begin{center}
\subfigure[ A plot of  $a'$ and $b'$ vrs. b for a=0.1]{\label{fig-aprbprap1}\includegraphics[height=70mm]{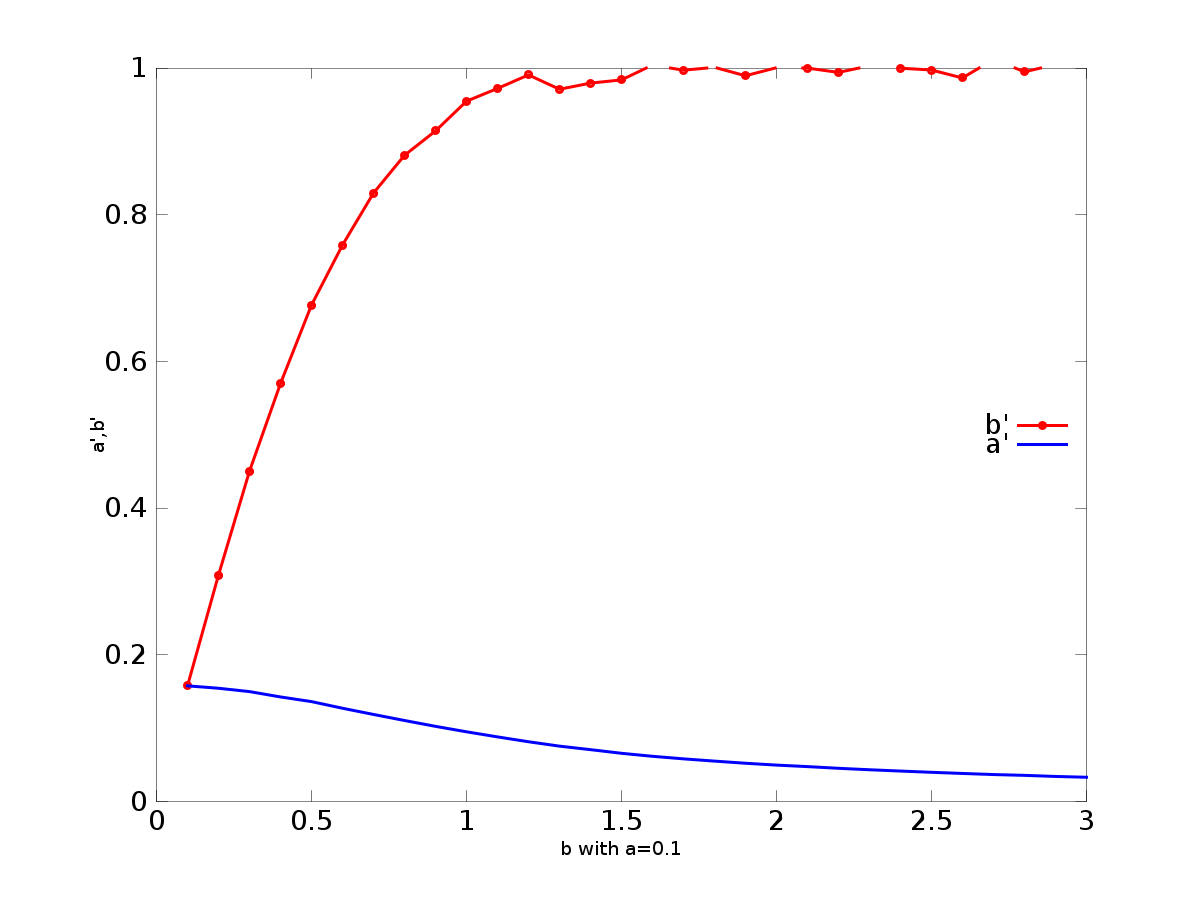}}
\subfigure[ A plot of $(1-8c_{f})$,  $(1-8c_{g})$(dotted line) vrs. b]{\label{disturasym}\includegraphics[height=70mm]{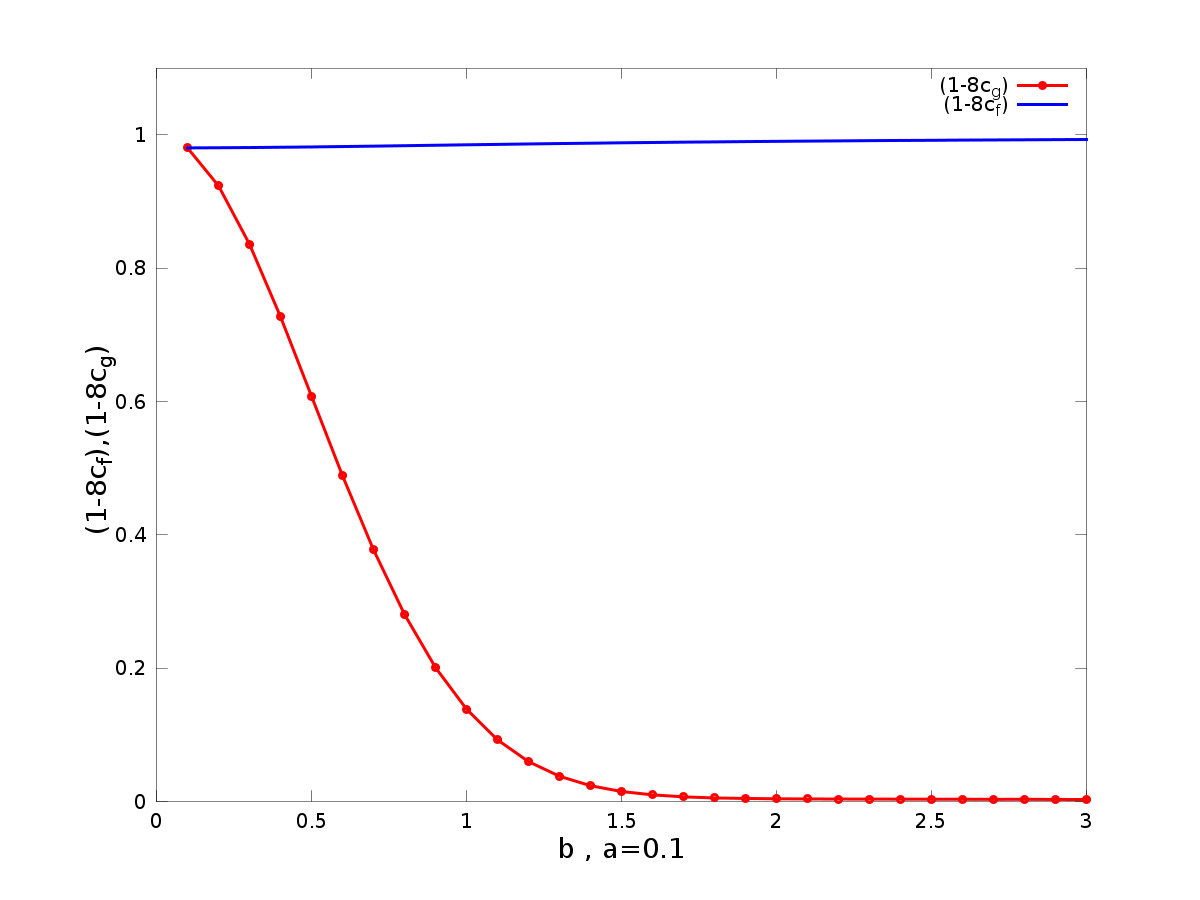}}
\end{center}
\caption{Effects of initial pointer states on the joint measurement and post-measurement system state for $b \geq a = 0.1$.}
\end{figure}

As seen in fig. \ref{disturasym} as the measurement becomes sharper ($b$ increases with $a=0.1$, $b'$tends to $1$ while $a'$ tends to $0$) , the $<\sigma_y>$ information 
of the initial density matrix  is 
almost kept intact,( ${\langle \sigma_y \rangle}_f = (1-8c_{g}) \langle \chi |\sigma_y |\chi \rangle $) while the  $<\sigma_x>$ information  
  gets very disturbed (${\langle \sigma_x \rangle}_f = (1-8c_{f}) \langle \chi |\sigma_x |\chi \rangle $) . This is reminiscent of a Heisenberg's Gammma-Ray microscope kind of a situation where
using a short wavelength light to reduce the uncertainty in position measurement of an electron disturbs the momentum of the electron .

b) Next we choose $a=1.0$.  The conclusions are almost similar with the exception that the joint measurement uncertainty (i.e the l.h.s of eqn. (\ref{eqajmuncer}) starts off at a much
 higher value compared to that of fig. \ref{fig-aprbprap1} (see fig. 3).

\begin{figure}
\label{fig-aprbpra1}
\begin{center}
\includegraphics[height=70mm]{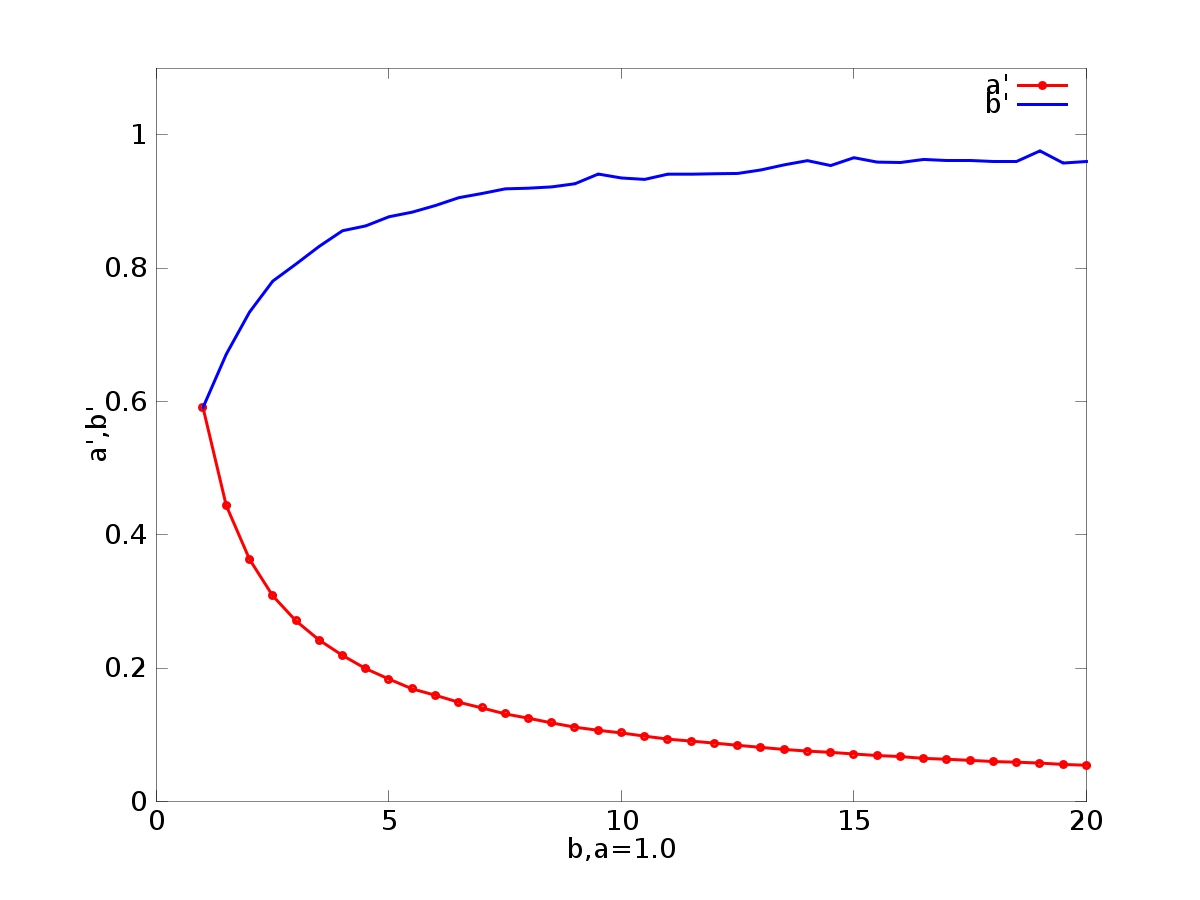}
\caption{A plot of $a'$ and $b'$ vrs. b, a=1.0 . Bottom curve shows $a'$ and top curve shows $b'$} 
\end{center}
\end{figure}

\section{Physics of the model and entanglement}
\label{sec-physics}
In this section we first try to understand the results obtained in the previous section. It is instructive to first look at a single approximate measurement 
arising from a von Neumann model. As mentioned before the situation is almost like a Stern-Gerlach Experiment.  
So we  consider a neutral particle of mass $m$ carrying spin-$\frac{1}{2}$ which propagating in the $z$-direction passes through  a magnetic field  $\vec{B}=-2B_0x \hat{x}$ 
for a time interval $\tau$. 
The position wavefunction of the particle before it enters the magnetic field is given by $\psi(x)= (\frac{1}{\sqrt{2\pi}}e^-{\frac{x^2}{2a^2}})^\frac{1}{2}$ .The interaction
Hamiltonian is taken to be,  $H=\frac{\vec{p}^2}{2m}  -\vec{S}.\vec{B}$.
The unitary evolution corresponding to this interaction is given by, 

\begin{equation}   
U= exp(iB_0 \tau x\otimes \sigma_x - i(\frac{p^2}{2m}\otimes 1_s)\tau) \mbox{  (in $\hbar=1$ \,units)}     
\end{equation}

The strong impulsive coupling approximation that we employed in section \ref{section-ajmortho} to neglect the kinetic energy term  amounts to $\frac{1}{2ma^2} \ll B_0 a$ . We assume that the particle is sufficiently
massive for this to hold for the values of a we have considered. We also note that in an usual Stern-Gerlach setting , initial position wavefunction is  
taken to be sharp and hence this assumption will breakdown for sufficiently small $a$ . After neglecting the kinetic energy we have,

\begin{eqnarray}
\label{sinfinstate}
|\psi_f \rangle &=& U (|\psi \rangle \otimes |\chi \rangle) \nonumber \\  
&=& \int_{p_x=-\infty }^{\infty}dp_x \, {}_x\langle + | \chi \rangle\psi(p_x - \lambda)  |p_x  \rangle\otimes |+\rangle_x  +  \int_{p_x=-\infty }^{\infty} dp_x \, {}_x \langle - | \chi \rangle\psi(p_x + \lambda) |p_x  \rangle\otimes |-\rangle_x 
\end{eqnarray}

with, $\lambda=B_0\tau $ and $\sigma_x |\pm \rangle_x = \pm 1 |\pm \rangle_x $ .

If $\lambda \gg \frac{1}{a}$ then we can distinguish between $ |+\rangle_x $ and  $ |-\rangle_x $ by looking at the momentum , as the 
Gaussian momentum distributions of width $\frac{1}{a} $ about $\pm\lambda$
do not essentially overlap.  The mean of the distribution moves to $\pm \lambda$ depending on whether the state is  $ |\pm\rangle_x $. If we take ($ p_x \geq 0 $ , $ p_x \leq 0$) 
to correspond to an unsharp measurement of $ \sigma_x $, then the POVM element $ G_+ $, characterising the unsharp measurement satisfies, 
\begin{eqnarray}
p(p_x\geq 0) &=& \langle \chi | G_+ |\chi \rangle \nonumber \\ 
&=&  \langle \chi| \frac{1}{2}(I + a'\sigma_1)| \chi \rangle  
\end{eqnarray}
with, $a' = 2F(\lambda a) - 1$ and  $ F(x)= \int_{-\infty}^{x} \frac{e^{-\frac{t^2}{2}}}{\sqrt{2\pi}}dt$.  Again as expected, as $\lambda a$ increases beyond $1$ , $F(\lambda a)$ and $a'$ moves
closer to one and the measurement becomes sharper. 

In the other limit , $\lambda \ll \frac{1}{a}$ distinguishability is lost and this is also reflected in $a'$ going to zero.     

Thus a necessary condition for an approximate measurement of $\vec{\sigma}.\hat{n}$ to be good can be taken to be its ability to distinguish between the  
eigenstates of  $\vec{\sigma}.\hat{n}$. Distinguishability in turn depends on the two length scales $\lambda$ , the distance by which the mean of the distribution moves
and the width of the distribution being $\frac{1}{a}$. This is also reflected in the fact that in the limit $\lambda \gg \frac{1}{a}$, eqn. (\ref{sinfinstate}) becomes 
a Schmidt decomposition and $|\psi_f \rangle $ becomes maximally entangled.

\subsection*{Joint measurement}
 The Arthur-Kelly model that we have considered can be thought to come from a magnetic field   $\vec{B}=-2B_0x \hat{x} -2B_0y \hat{y}$ ($\lambda=B_0\tau $ has been taken                               
to be 1). As mentioned before the non-zero divergence of this field is not a serious issue. We could also have taken   $\vec{B}=-2B_0x \hat{y} -2B_0y \hat{z}$  and measured 
$p_x, p_y$ in order to have an approximate joint measurement of $\sigma_y$ and  $\sigma_z$ .
        
 After the measurement of momentum , the probability of obtaining $p_x$ is given by (see eqn.(\ref{probability}))
\begin{equation}
p(p_x) = \int_{-\infty}^{\infty} (|e^0|^2 + |f^0|^2 + |g^0|^2)dp_y + 2  \int_{-\infty}^{\infty} e^0{f^0}^* dp_y  \langle \chi | \sigma_x | \chi \rangle
\end{equation}
  
and
\begin{equation}
\label{eqp1geq0}
p(p_x \geq 0) = \frac{1}{2} + \frac{a'}{2}  \langle \chi | \sigma_x | \chi \rangle 
\end{equation}
with
\begin{equation}
 a'  = \int_{p_x=0}^{+\infty} \int_{p_y=-\infty}^{+\infty} 4(f^0{e^0}) dp_xdp_y  .
\end{equation}

          The origin of complementarity between $\sigma_x$ and  $\sigma_y$ in this model  can be understood from the way the effective length scales governed by the movement 
of the mean momenta $<p_x>$  and $<p_y>$ change as we make one of the initial momentum wavefunctions sharper or broader. The Ehrenfest Theorem applied on the Hamiltonian (\ref{model})
gives,
\begin{eqnarray}
\label{eq-Ehrenfest}
\dot{\langle p_x \rangle} &=&  \langle \sigma_x \rangle ,\nonumber \\
\dot{\langle p_y \rangle } &=&   \langle \sigma_y \rangle .
\end{eqnarray}

Now as  we saw in the section \ref{subsection-rdm} and  \ref{subsection-distur} ,the  effect of the interaction (\ref{unitary}) on the spin state of the system is that of an asymmetric depolarising 
channel that disturbs the $\langle \sigma_x \rangle$ component of the density matrix while keeps  the  $\langle \sigma_y \rangle$ almost intact as the initial $y$ momentum wavefunction
is made much sharper than the $x$ momentum one. Eq. (\ref{eq-Ehrenfest}) shows that rate of change of average momentum in the x and y directions is similarly affected. Thus  with the increasing
sharpness of the  initial $y$-momentum wavefunction, the movement of the $p_y$ mean is not affected much , but the $p_x$  mean moves very little. It becomes harder
to distinguish  between  $\sigma_x$ eigenstates by looking at $p_x$ distribution after the interaction and  unsharpness of  $\sigma_x$ measurement increases.  

\begin{figure}
\begin{center}
\subfigure[A plot of $P(p_1)$ and $P1(p_1)$ for a=5.0 , b=1.0]{\label{fig-a5b1}\includegraphics[height=70mm]{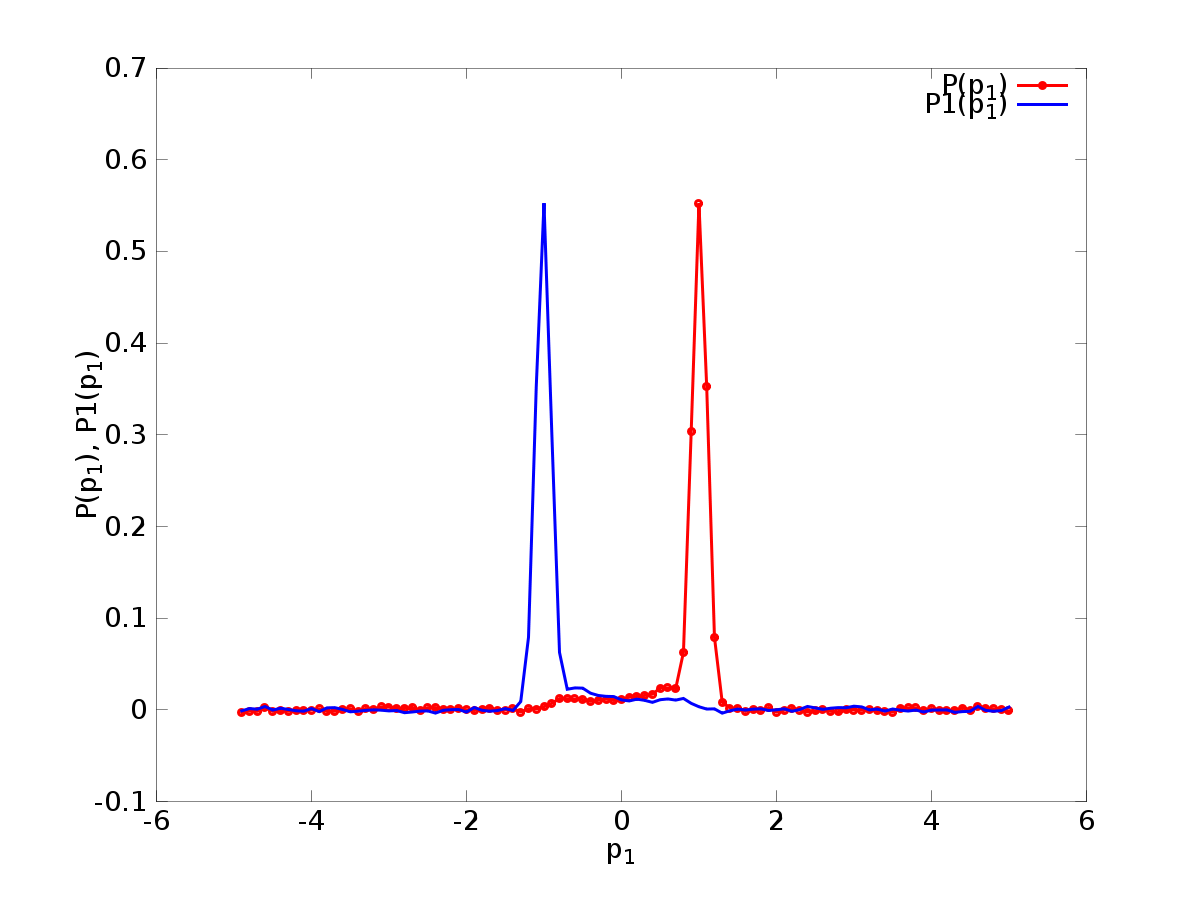}}
\subfigure[A plot of $P(p_1)$ and $P1(p_1)$ for a=5.0 , b=25.0]{\label{fig-a5b25}\includegraphics[height=70mm]{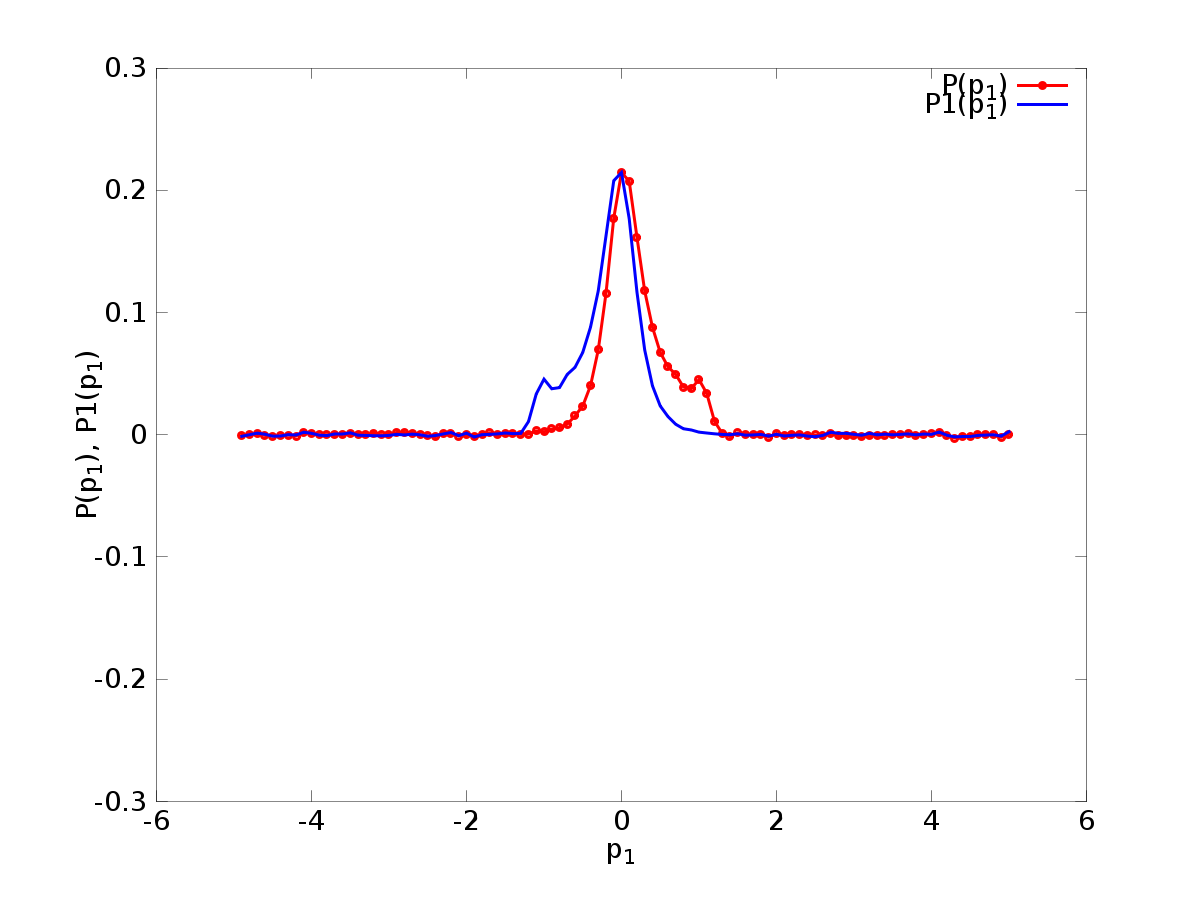}}
\end{center}
\caption{A plot of probability of measuring $p_1$ after pre-measurement with $p_1$. $P(p_1)$  represents probability of $p_1$
for initial spin states $|+\rangle_x$ and $P1(p_1)$ represents that for initial spin state $|-\rangle_x$ . }
\label{figPp1}
\end{figure}

This is illustrated in fig. \ref{figPp1} .

\paragraph{Disturbance due to the measurement:}
In order to understand  the asymmetric depolarising action of the interaction on the spin state we look at the rdm of the system after the interaction once more.
\begin{eqnarray}
\rho_f^s &=&  Tr_{1,2} (|\psi_f \rangle \langle \psi_f |) \nonumber \\
&=& Tr_{1,2}(e^{i(x_{op}\otimes \sigma_x +y_{op} \otimes \sigma_y)} |\psi \rangle \langle \psi | \otimes |\chi \rangle \langle \chi | e^{-i(x_{op}\otimes \sigma_x +y_{op} \otimes \sigma_y)})\nonumber \\
&=& \int_{x,y=-\infty}^{+\infty} e^{i(x \sigma_x +y  \sigma_y)} |\chi \rangle \langle \chi| e^{-i(x \sigma_x +y  \sigma_y)} |\psi_1(x)|^2 |\psi_2(y)|^2 dxdy \nonumber \\ 
\label{eq-Bloch_rot}
&=& \int_{x,y=-\infty}^{+\infty} e^{i\frac{\vec{\sigma}.\hat{r}}{2} (2r) } |\chi \rangle \langle \chi| e^{-i\frac{\vec{\sigma}.\hat{r}}{2} (2r) } \:\frac{e^{-(\frac{x^2}{a^2} +\frac{y^2}{b^2}})}{2 \pi ab }dxdy 
\end{eqnarray}
 
with $r=\sqrt{x^2 + y^2}$, $\hat{r}$ denoting the radius and the unit radial vector respectively in polar coordinates .
As eqn. (\ref{eq-Bloch_rot}) shows, the rdm of the spin part of the system  after the interaction is a mixture of rotated states about $-\hat{r}$ by an angle $(2r \mbox{ mod } 2\pi). $ 
The weight of a rotated state about $-\hat{r}$ in the mixture is the Gaussian  probability density of the initial position of the particle. Thus as we increase $b$ keeping $a$ fixed ,
the weight of states which are rotated near about the $y$ axis increases.   Hence the $\langle \sigma_y \rangle$ component of the initial density matrix is disturbed more and more while the  
$\langle \sigma_x \rangle$ component is almost kept intact. 
  
                We earlier saw that the  the quality of unsharpness $a'$ increase and show a maximum in the symmetric case (fig. \ref{fig-a'}). As we saw the disturbance  
due to the measurement governed by the plot of $(1-8c_{f})$ vrs. $a$ , also show a minimum. The maximum in the $a'$ vrs. a curve was also there in \cite{kienzler}. 

                   We argue that the maxima is due to the $2r$ factor in eq.(\ref{eq-Bloch_rot}) . For smaller values of $a$ the rotations are  constrained to smaller angles. 
 To understand this we take a magnetic field of the form $\vec{B}=-2B_0 \hat{r} $.
This removes the $r$ factor in \ref{eq-Bloch_rot} . $a'$ is then seen to  changes very little with a (from 0.27 to .285) . The disturbance characterised by $c_{f}$ remains constant
at .088 to about four orders of magnitude. The slight increase in $a'$ is presumably due to the modification of the Ehrenfest eqns. (\ref{eq-Ehrenfest}) due to the non-linearity
in the magnetic field.  

\section*{ Entanglement between detectors and system}

We next consider the entanglement between the system and the two detectors. The detectors are infinite dimensional while the system of course is two-dimensional.
Though strictly speaking only correlations between the detectors and system is required for the measurement on the detectors to reflect  measurement statistics of the system , one expects 
entanglement to play a role in this kind of a scenario.    

\subsection{Entanglement between the joint detector system and the system}
First, we consider the entanglement between the two detectors (considered as a single system) and the qubit system. As the final state $|{\psi}_f \rangle$ of 
the system and the detectors after pre-measurement is pure
this entanglement is simply given by the von Neumann entropy of the reduced density of the system after the detectors are traced out. We have,
\begin{equation}
|{\psi}_f \rangle = \int dp_1 dp_2 e^0 |p_1,p_2 \rangle \otimes |\chi \rangle +  \int f^0 dp_1 dp_2 |p_1,p_2 \rangle \otimes \sigma_x |\chi \rangle + \int  dp_1 dp_2 g^0 |p_1,p_2 \rangle \otimes \sigma_y |\chi \rangle \nonumber
\end{equation}

Let, 
$\rho_f^s = Tr_{12} |\psi_f \rangle \langle \psi_f |  $. We have,
\begin{eqnarray}
 \rho_f^s =& \int_{p_1 =-\infty}^{\infty} \int_{p_2 =-\infty}^{\infty} |e^0|^2 dp_1 \, dp_2 \,|\chi \rangle \langle \chi | + \nonumber \\ 
& \int_{p_1 =-\infty}^{\infty} \int_{p_2 =-\infty}^{\infty}  |f^0|^2 dp_1 \, dp_2 \,(\sigma_x |\chi \rangle \langle \chi | \sigma_x) + \nonumber \\  
& \int_{p_1 =-\infty}^{\infty} \int_{p_2 =-\infty}^{\infty}  |g^0|^2 dp_1 \, dp_2 \,(\sigma_y |\chi \rangle \langle \chi | \sigma_y) 
\end{eqnarray}

Now, for $|\chi \rangle \langle \chi |= \frac{1}{2}(I + x\sigma_1 + y\sigma_2 + z\sigma_3) $ we have ,
\begin{equation}
 \rho_f^s = \frac{I}{2} + \frac{\sigma_x}{2}(x + 4c_{ff^*}(1-x) - 4xc_{gg^*}) + \frac{\sigma_y}{2}(y + 4c_{gg^*}(1-y) - 4yc_{ff^*}) +  \frac{\sigma_z}{2}(1-4(c_{ff^*} + c_{gg^*} ))    
\end{equation}

\begin {figure}
\begin{center}
\subfigure[A plot of $S(\rho_f^s)$ vrs. a=b]{\label{fig-ent1}\includegraphics[height=70mm]{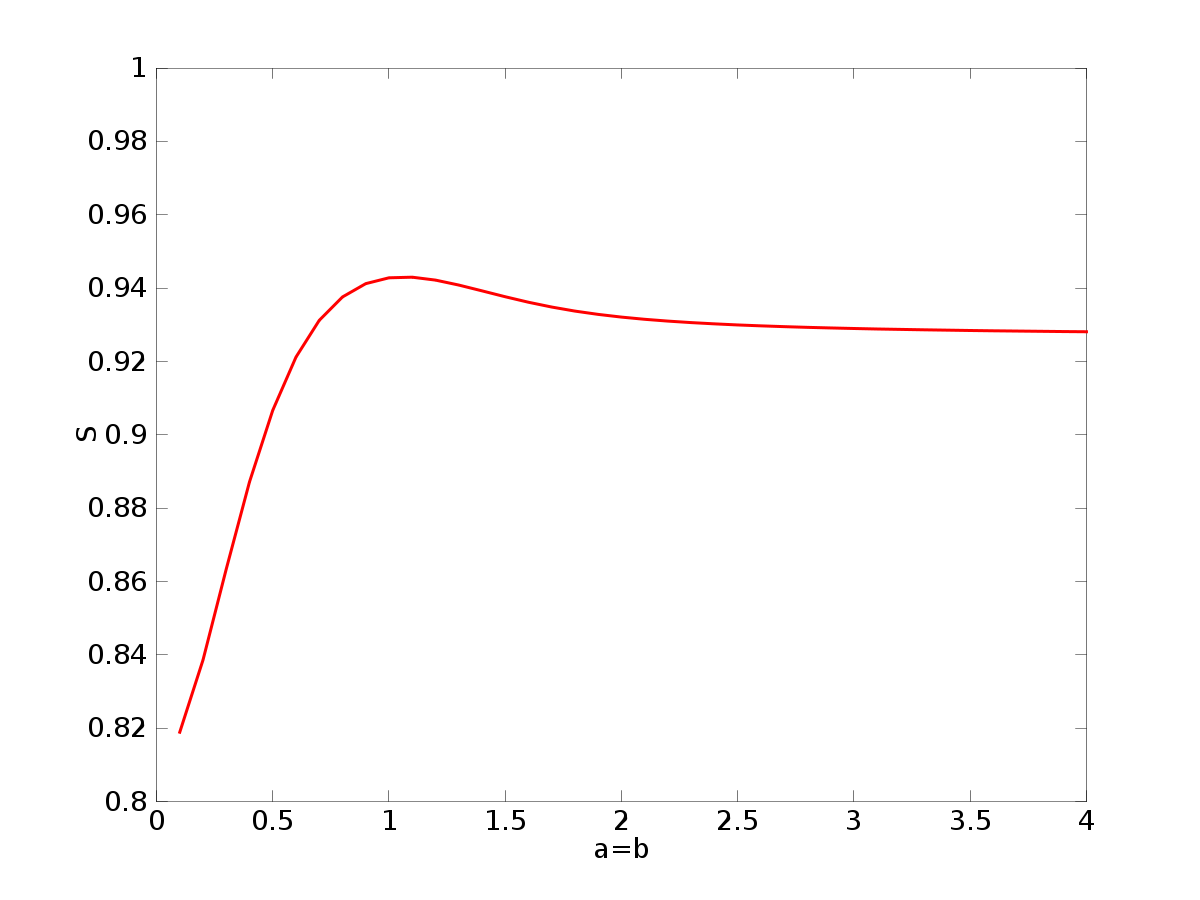}} 
\subfigure[A plot of $S(\rho_f^s)$ vrs. b for a=0.1]{\label{fig-ent2}\includegraphics[height=70mm]{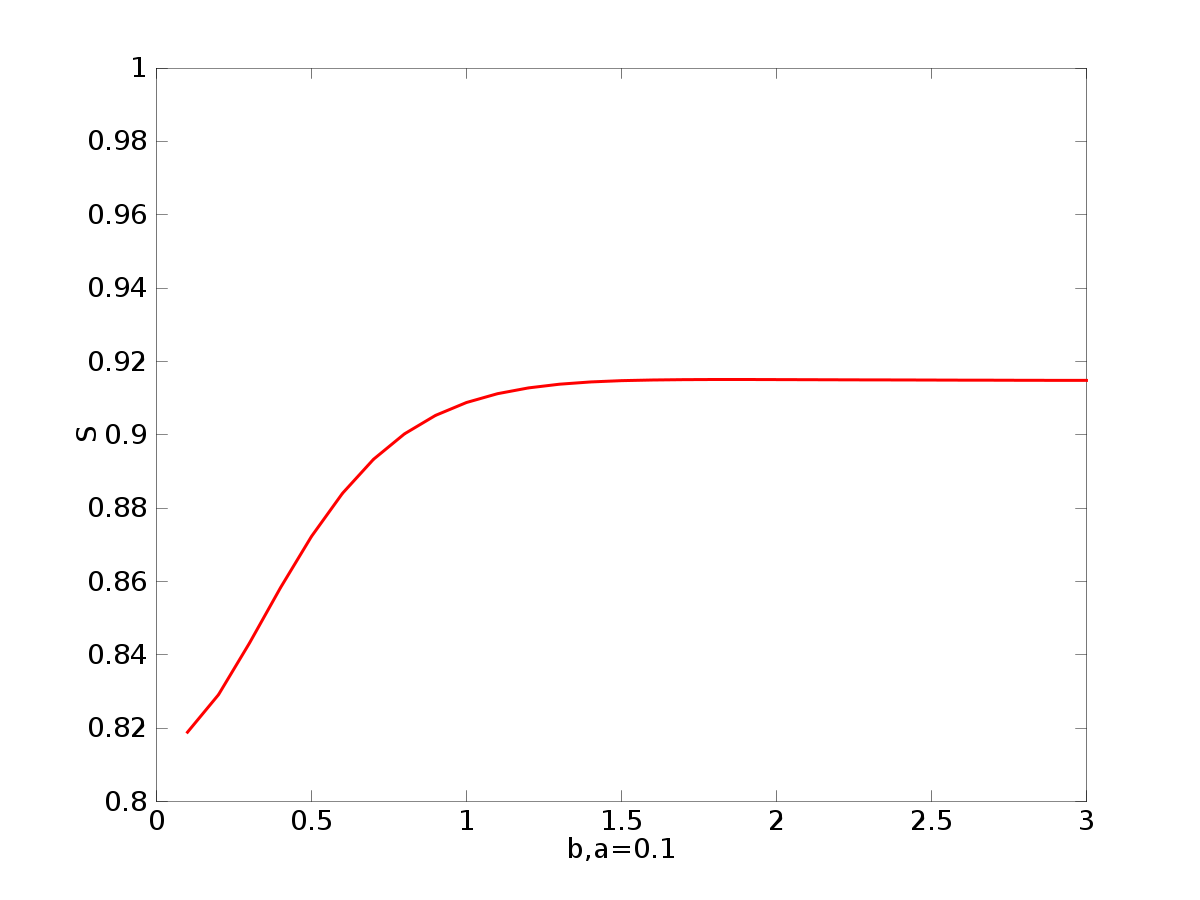}}
\end{center}
\caption{ Entanglement between the joint detector system and the qubit system as reflected by the Von Neumann entropy of the r.d.m of the system after the configuration degrees are traced out.}
\label{entjointdetsys}
\end{figure}

Taking $x,y= \frac{1}{2}$ and $z=0$ the von Neumann entropy of the system is given by fig \ref{entjointdetsys},

The basic feature is that when the sharpnesses of the initial momentum wavefunctions are low the entanglement is relatively low , increasing as the sharpness increases.  In both the symmetric and the asymmetric cases 
maximal entanglement is not reached.

\subsection*{Entanglement between detector and system}
In this subsection we try to see the entanglement in the mixed state of one of the detectors and the system after the other detector is traced out. Now the situation is 
made difficult to handle by the fact
that the detector is an infinite dimensional configuration degree of freedom . For this reason we project the 
state of the detector into two dimensional momentum subspaces.  The average entanglement considering all such outcomes gives  a lower bound on the entanglement of the $\infty \times 2$ state of the system and detector 
as projection being a local action cannot increase the the entanglement on the average.

\begin{figure}
\begin{center}
\subfigure[ A plot of  $P(p_1)$ , $P1(p_1)$ vrs. $p_1$ for $a=b=0.1$]{\label{fig-distrosymp1}\includegraphics[height=70mm]{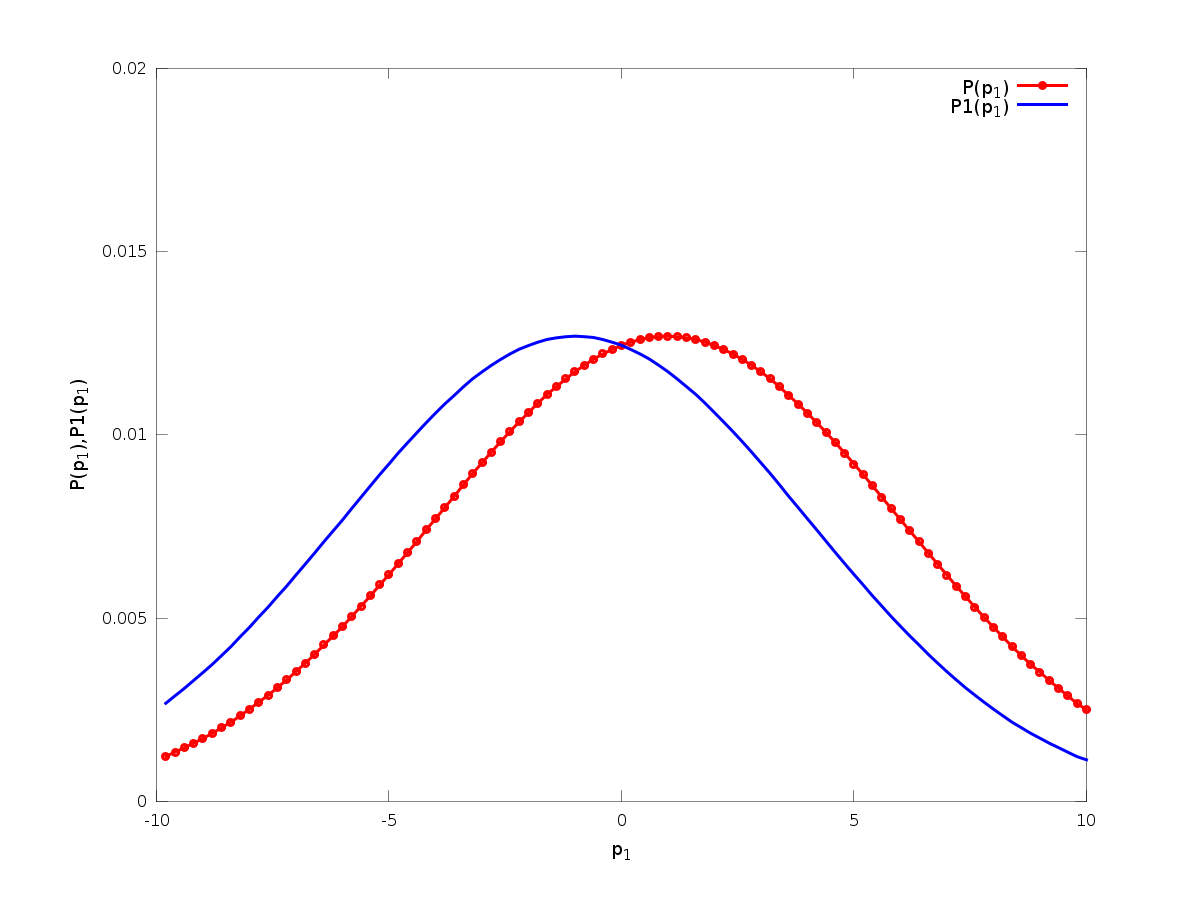}} 
\subfigure[A plot of  $P(p_1)$ , $P1(p_1)$ vrs. $p_1$ for $a=0.5 , b=0.1$]{\label{fig-ent2}\includegraphics[height=70mm]{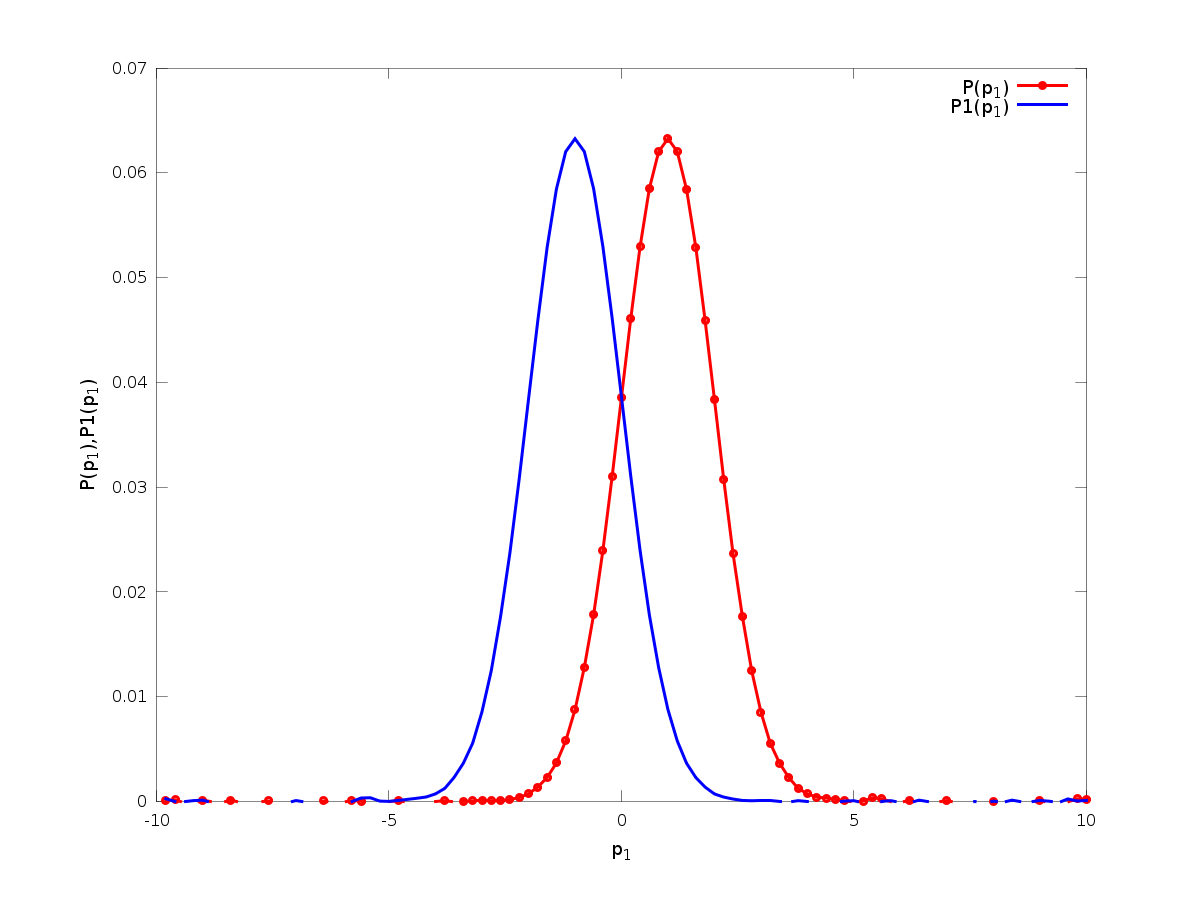}}
\end{center}
\caption{Probability distribution for obtaining  $p_1$ .  $P(p_1)$ and $P1(p_1)$ denotes respectively probability of obtaining $p_1$ for initial system state $|+ \rangle_x$ and 
$|- \rangle_x$. } 
\label{fig-probp1plot}
\end{figure}

In fig. \ref{fig-probp1plot} we clearly see that the probability in eq. \ref{eqp1geq0} becomes more and more greater than 0.5 as a increases with respect 
to b (for a $|+\rangle_x$ spin state), signifying an increase in $a'$ as well. Depending on whether the state is  $|+\rangle_x$ or  $|-\rangle_x$ the probability peaks 
around $p_x=\pm 1$ (as the initial distribution is symmetric and we have taken the time of interaction 
to be unit , this follows from eq. (\ref{eq-Ehrenfest}) for almost sharp measurements ). Thus  
correlation is likely to be highest  between the system and the detector in the two dimensional momentum subspace around $p_x = \pm 1.0 $ . 

As a measure of entanglement we consider concurrence which is defined below. We have considered projections in different two-dimensional momentum subspaces. The concurrence    
for  states projected into different subspaces is qualitatively seen to follow the same behavior as state projected into $p_1 =\pm 1$ subspace. However for $a$ much
greater than $b$ , numerics becomes difficult when we consider projections into subspaces far from $\pm 1$. This is  due to momentum distribution peaking around $p_1=\pm1$ 
(see fig. \ref{fig-probp1plot} ) and we expect the entanglement to fall for $a$ much greater than $b$ in these subspaces. Also, as the probability of obtaining 
such projections  falls  
we do not expect the concurrence of states projected into momentum subspaces far from $\pm1$ to contribute much to the average concurrence. 

Thus we take the concurrence of the projected mixed state into $p_1=\pm 1$ subspace to be an indicator of how the average concurrence for all possible projections
on to two-dimensional subspaces behave ( the first concurrence multiplied by probability of obtaining $p_1=\pm1$ is of course a lower bound for the average concurrence) as we increase 
sharpness of one initial detector momentum wavefunction ($a$) keeping the other fixed ($b$) (eqns. \ref{inistate1} and \ref{inistate2}).

We consider  the spin state to be the symmetric state $|\chi \rangle \langle \chi | = \frac{1}{2} (I + \frac{\sigma_x}{\sqrt2} +  \frac{\sigma_y}{\sqrt2})$. 

Let us first consider the entanglement between the first detector and the system . Let $\rho_f^{1} = Tr_2(|\psi_f \rangle \langle \psi_f |)$ . ( $|\psi_f \rangle$ is the final state of the system and the detectors after
the measurement interaction.). Because of ``parity properties`` we have ,
\begin{eqnarray} 
\int_{-\infty}^{\infty} e^0(1,p_2)e^0(-1,p_2) dp_2 &=& \int_{-\infty}^{\infty} e^0(1,p_2)e^0(1,p_2) dp_2 := {E^0}(1) ,\nonumber \\
\int_{-\infty}^{\infty} f^0(1,p_2)f^0(1,p_2) dp_2 &=& - \int_{-\infty}^{\infty} f^0(1,p_2)f^0(-1,p_2) dp_2 := {F^0}(1), \nonumber \\
\int_{-\infty}^{\infty} g^0(1,p_2)g^0(1,p_2) dp_2 &=&  \int_{-\infty}^{\infty} g^0(1,p_2)g^0(-1,p_2) dp_2 := {G^0}(1) ,\nonumber \\
\int_{-\infty}^{\infty} e^0(1,p_2)f^0(1,p_2) dp_2 &=& - \int_{-\infty}^{\infty} e^0(1,p_2)f^0(-1,p_2) dp_2 := E^0F^0(1) .
\end{eqnarray}

Let, $ \textbf{P} = |p_1 =1 \rangle \langle p_1 =1 | +  |p_1 =-1 \rangle \langle p_1 =- 1 | $. We have,
\begin{eqnarray}
\rho^1 = \textbf{P} \rho_f^{p_1}\textbf{P} \nonumber \\
 & = {E^0}(1) (|1 \rangle \langle 1| + |1 \rangle \langle -1| + |-1 \rangle \langle 1| +  |-1 \rangle \langle -1|) \otimes |\chi \rangle \langle \chi | + \nonumber \\
& +  {F^0}(1) ( |1 \rangle \langle 1| -   |-1 \rangle \langle 1| -  |1 \rangle \langle -1| +  |-1 \rangle \langle -1|) \otimes \sigma_1 |\chi \rangle \langle \chi | \sigma_1 \nonumber \\  
& + {G^0}(1)  (|1 \rangle \langle 1| + |1 \rangle \langle -1| + |-1 \rangle \langle 1| +  |-1 \rangle \langle -1|)\otimes \sigma_2 |\chi \rangle \langle \chi | \sigma_2 \nonumber \\ 
& + (E^0F^0(1)|1 \rangle \langle 1| + E^0F^0(1)|-1 \rangle \langle 1| - E^0F^0(1)|1 \rangle \langle -1| - E^0F^0(1)|-1 \rangle \langle -1| ) \otimes |\chi \rangle \langle \chi| \sigma_1 \nonumber \\  
& + (E^0F^0(1)|1 \rangle \langle 1| - E^0F^0(1)|-1 \rangle \langle 1| + E^0F^0(1)|1 \rangle \langle -1| - E^0F^0(1)|-1 \rangle \langle -1| ) \otimes \sigma_1 |\chi \rangle \langle \chi |, 
\end{eqnarray}
where $|1\rangle \equiv | p_1 =1 \rangle$ and $|-1\rangle \equiv | p_1 =-1 \rangle$ .
Now for the spin state we have considered , we have the the unnormalized density matrix $\rho^1$ in the 
basis 

$|1\rangle \equiv | p_1=1 \rangle \otimes |0 \rangle$ , 
$ |2\rangle \equiv |p_1=1\rangle \otimes |1 \rangle$, $ |3\rangle \equiv |p_1=-1\rangle \otimes |0 \rangle$ and $ |4\rangle \equiv |p_1=-1\rangle \otimes |1 \rangle$  given by, 
\begin{eqnarray}
\rho^1_{11} &=& 0.5({E^0}(1) + {F^0}(1) + {G^0}(1)) + 0.707E^0F^0(1) \nonumber \\  
\rho^1_{12} &=&  0.5({E^0}(1) - {F^0}(1) + {G^0}(1)) + 0.707iE^0F^0(1) \nonumber \\
\rho^1_{13} &=& 0.353(1-i) {E^0}(1) + 0.353 (1+i) {F^0}(1) - 0.353 (1+i){G^0}(1) +  E^0F^0(1) \nonumber \\  
\rho^1_{14} &=& 0.353(1-i) {E^0}(1) -  0.353 (1+i) {F^0}(1)  - 0.353 (1+i){G^0}(1) \nonumber \\
\rho^1_{22} &=& 0.5({E^0}(1) + {F^0}(1) + {G^0}(1)) - 0.707E^0F^0(1) \nonumber \\ 
\rho^1_{23} &=& 0.353(1-i){E^0}(1) - 0.353(1+i){F^0}(1) - 0.353(1+i){G^0}(1) \nonumber \\
\rho^1_{24} &=& 0.353(1-i){E^0}(1) - E^0F^0(1) + 0.353(1+i){F^0}(1) - 0.353(1+i){G^0}(1) \nonumber \\
\rho^1_{33} &=& 0.5({E^0}(1) + {F^0}(1) + {G^0}(1)) + 0.707E^0F^0(1) \nonumber \\
\rho^1_{34} &=& 0.5({E^0}(1) - {F^0}(1) + {G^0}(1)) - 0.707i E^0F^0(1) \nonumber \\
\rho^1_{44} &=&  0.5({E^0}(1) + {F^0}(1) + {G^0}(1)) - 0.707 E^0F^0(1) 
\end{eqnarray}

with $Tr(\rho_1) = 2({E^0}(1) + {F^0}(1) + {G^0}(1)) $

After normalisation we apply the Peres-Horodecki PPT criterion to check for entanglement (\cite{peres}) .  As a measurement of entanglement we use the concurrence  which for a 2x2 density matrix
is defined as $C(\rho)= \mbox{ max }(0, \lambda_1 - \lambda_2 - \lambda_3  - \lambda_4 ) $ , where $\lambda_i$ denote in decreasing order the square root of the 
eigenvalues of the non-Hermitian matrix  $\rho \tilde{\rho} $ ,with $\tilde{\rho} = (\sigma_{y}\otimes\sigma_{y})\rho^{*}(\sigma_{y} \otimes \sigma_{y}) $.(\cite{wooters})

\begin{figure}
\begin{center} 
\includegraphics[height=70mm]{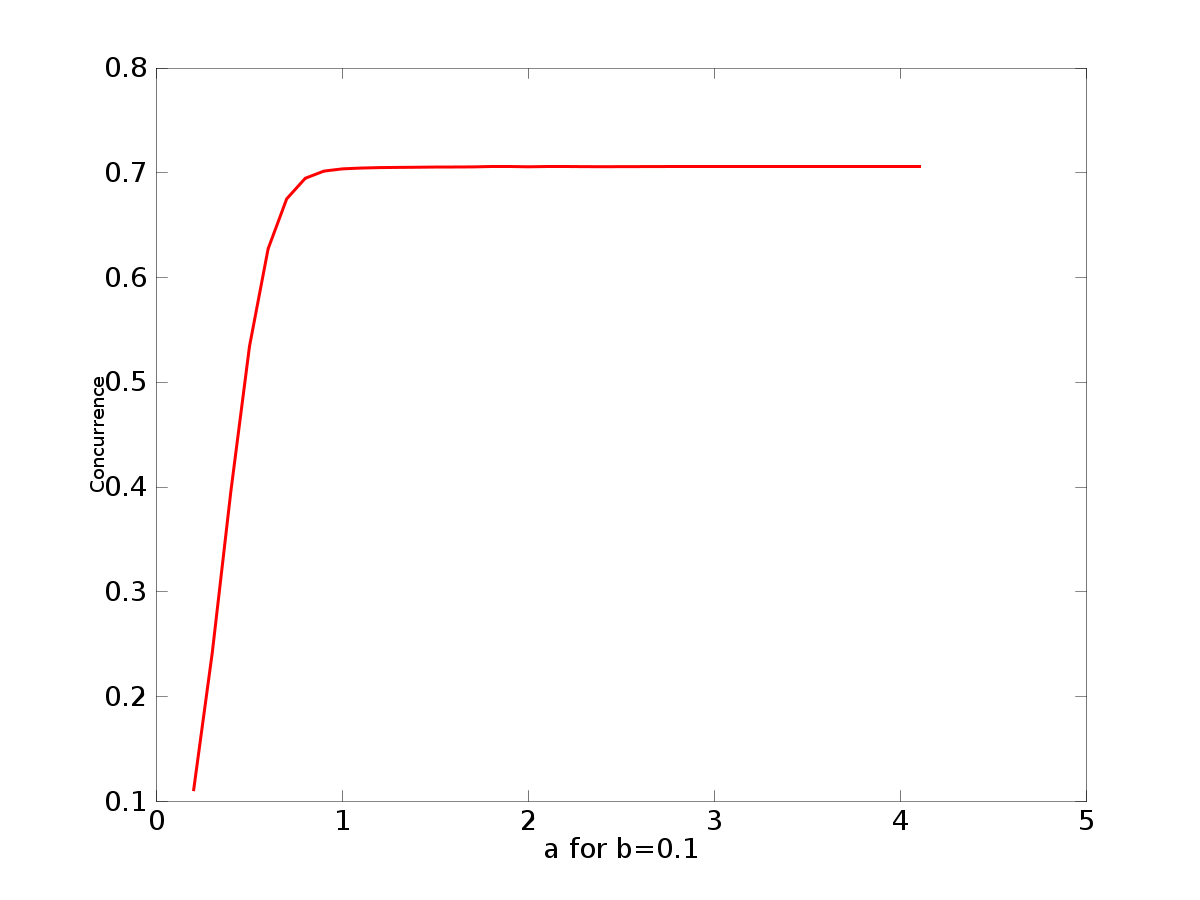}
\end{center}
\caption{Concurrence of the rdm of system and first detector after projection into the subspace $p_1=\pm1$ vrs. a for b=0.1}
\label{concurrence}
\end{figure}

In fig. \ref{concurrence} we see that the concurrence increases as $a$ increases with fixed $b$, i,e as the $\sigma_x$ measurement becomes sharper . We expect the entanglement between
the system and the detector to behave similarly.

\section{Effect of the symmetries of the underlying Hamiltonian on the POVM elements}
\label{section-povmsym}
It was shown in \cite{Myunck} in the context of a Stern-Gerlach type Hamiltonian how the symmetries of the underlying Hamiltonian like the one given in eqn. ( \ref{model} ) can be used to do some particular 
unsharp measurement. The various cases considered there can be seen to follow from the following lemma. 

Consider an Arthur-Kelly like measurement process and after the  measurement on the detectors is made.

\textbf{Lemma 1:}  Let $H$ be  an Arthur-Kelly like Hamiltonian of the form $H= f(q_1, q_2) \otimes \sigma_x + g(q_1, q_2) \otimes \sigma_y$ , which has a symmetry given by  $[A \otimes B, H] = 0$ . Here $A$ and $B$ 
are unitaries acting respectively on the joint detector space  and the spin space. Let $ P_{|\chi \rangle} (p_1, p_2) $  represent the probability of obtaining the momenta values $p_1$ , $p_2$ if the system 
is initially in  the state $|\chi \rangle$ .  Further let the initial joint detector state $|\psi \rangle$ have the symmetry $A$ 
so that $A|\psi \rangle = e^{i\phi}|\psi \rangle$. 
Then for  the initial system state  $|\chi\rangle$ and a new basis in the joint detector space  $|p'_1, p'_2 \rangle = A |p_1, p_2 \rangle$ , we have 
\begin{equation}
\label{povmsym}
  P_{|\chi \rangle} (p_1, p_2) =  P_{B|\chi \rangle} (p'_1, p'_2) .
\end{equation}

\paragraph{Proof:}   

\begin{equation}
\label{firsteq}
P_{|\chi \rangle} (p_1, p_2) = Tr_{1,2}[\mathbf{P}[|p_1, p_2 \rangle]Tr_s(e^{-iH} \mathbf{P} [|\psi \rangle \otimes |\chi \rangle] e^{iH})] 
\end{equation}
with $\mathbf{P}[|\eta\rangle]$ denoting projector on $|\eta\rangle$.

Now replacing $e^{-iH}$ by $(A^{\dagger} \otimes  B^{\dagger})e^{-iH} (A \otimes B$) in (\ref{firsteq}) and using $A|\psi \rangle = e^{i\phi}|\psi \rangle$    we have ,
\begin{equation}
\label{secondeq}
 P_{|\chi \rangle} (p_1, p_2) = Tr_{1,2}[\mathbf{P}[|p_1, p_2 \rangle]Tr_s((A^{\dagger} \otimes  B^{\dagger})e^{-iH}(|\psi\rangle \langle \psi | \otimes B |\chi \rangle \langle \chi |B^{\dagger}) e^{iH}(A \otimes B)] 
\end{equation}

Let ,  
\begin{equation}
\label{step1}
 e^{-iH}(|\psi\rangle \langle \psi | \otimes B |\chi \rangle \langle \chi |B^{\dagger})e^{iH} = e^{-iH}(\mathbf{P} [|\psi \rangle \otimes B|\chi \rangle]) e^{iH}=\sum_j C_j \otimes D_j , 
\end{equation}
 
where the index $j$ runs over a countable set.

$C_j$ s and $D_j$s  are operators acting respectively on the joint Hilbert space of the two detectors and the system Hilbert space.  

Using $B^{\dagger}B= I$ we  have  from eqns. (\ref{firsteq}) , (\ref{secondeq}) and (\ref{step1}):  
\begin{equation}
Tr_s(e^{-iH}\mathbf{P}[|\psi \rangle \otimes |\chi \rangle]e^{iH}) = \sum_j Tr_s(D_j)(A^{\dagger}C_j A) .   
\end{equation}

So,
\begin{eqnarray}
 P_{|\chi \rangle} (p_1, p_2) &= &\sum_j Tr_{1,2}[|p_1,p_2 \rangle \langle p_1,p_2 |Tr_s(D_j)(A^{\dagger}C_j A)] \nonumber \\ 
& = & \sum_j \langle p_1, p_2 |A^{\dagger}C_j A | p_1, p_2 \rangle Tr_s(D_j) \nonumber \\ 
& = &  \sum_j \langle p'_1, p'_2 |C_j |p'_1, p'_2 \rangle Tr_s(D_j) 
\end{eqnarray}
Again,
\begin{eqnarray}
P_{B|\chi \rangle} (p'_1, p'_2) &=& Tr_{1,2}[\mathbf{P}[|p'_1, p'_2 \rangle]Tr_s(e^{-iH} \mathbf{P} [|\psi \rangle \otimes B|\chi \rangle] e^{iH})] \nonumber \\
&=& Tr_{1,2}[\mathbf{P}[|p'_1, p'_2 \rangle]Tr_s(\sum_j C_j \otimes D_j)], (\mbox{from eqn. (\ref{step1})})  \nonumber \\
&=&  \sum_j \langle p'_1, p'_2 |C_j |p'_1, p'_2 \rangle Tr_s(D_j)  
\end{eqnarray}
\hfill $\square$ 

\section{Approximate joint measurement in arbitrary directions}
Let $I_i$ (i=1,2) denote the operation of reflection in the detector space about $q_i$-axis (where $q_1=x$ and $q_2=y$). Consider a Hamiltonian of the entire system (i.e, two detectors plus the qubit)
which satisfies $[I_1 \otimes \sigma_x, H]=0$ and an initial joint state 
$\psi$ of the two detectors jointly, that satisfies $I_1|\psi \rangle= |\psi \rangle$ . Then eqn. (\ref{povmsym}) yields, (with $E$ denoting the POVM element for a particular probability)
\begin{eqnarray}
 P_{|\chi \rangle} (p_1, p_2) &=& P_{\sigma_x|\chi \rangle} (p_1, -p_2) . \nonumber\\
\mbox{So,   }\;\langle \chi|E(p_1,p_2)|\chi \rangle & = & \langle\chi|\sigma_x E(p_1,-p_2)\sigma_x |\chi\rangle \nonumber\\
\mbox{i.e   }\; E(p_1, p_2) & = &\sigma_x E(p_1,-p_2)\sigma_x \nonumber\\
\mbox{and integrating over $p_2$,    }\;   [E(p_1), \sigma_x]&=&  0 . \nonumber\\ 
\mbox{So we can write,  } \;E(p_1) & = & \frac{1}{2}(\alpha((p_1) I + \beta((p_1) \sigma_x) \nonumber \\
\mbox{where $\alpha(p_1)$ and $\beta(p_1)$ are real nos.} \nonumber \\
\label{povmform}
\mbox{Hence we have,   }\; E(p_1 \geq 0) & = & \frac{1}{2} (\alpha'I + \beta'\sigma_x)
\end{eqnarray}
with constant $\alpha'$ and $\beta'$. 

As the Hamiltonian $H$ of eqn. (\ref{model}) satisfies  $[I_1 \otimes \sigma_x, H]=0$ and as the initial joint detector state $|\psi\rangle =|\psi_1\rangle \otimes |\psi_2\rangle$, where $|\psi_1\rangle$
and  $|\psi_2\rangle$ are given respectively by eqns. (\ref{inistate1}) and (\ref{inistate2}) satisfies  both  $I_1|\psi \rangle= |\psi \rangle$ and  $I_2|\psi \rangle= |\psi \rangle$ , it is clear from 
eqn. \ref{povmform} that how the approximate joint measurement of  $\sigma_x$ and $\sigma_y$ with the marginals given by eqn. (\ref{povm-marginal})   arises out of the measurement of $p_1$ and $p_2$.  

Further consider an $H$ which in addition to having the symmetry mentioned in the beginning of the present section, has a further rotational symmetry: 
 $[R(\theta) \otimes S_z(\theta), H]= 0$, where $R(\theta)$ and $S_z(\theta)$ respectively denote  rotation by an angle $\theta$ in the detector Hilbert space
(i.e, the Hilbert space operation corresponding to rotation in the $q_1$, $q_2$ plane)  and the spin space (about z-axis).      
From eqn. (\ref{povmsym}) we have,
\begin{equation}
\label{pov-sym-rot}
 P_{|\chi \rangle} (p_1, p_2) =  P_{S_z(\theta)|\chi \rangle} (p'_1, p'_2) 
\end{equation}
with ${(p'_1, p'_2)}^T = R(\theta){(p_1, p_2)}^T$

Integrating both sides over the region   $p_1 \geq 0$, $ -\infty \leq p_2 \leq \infty $   we have of eqn. (\ref{pov-sym-rot}), 

\begin{eqnarray}
P_{|\chi \rangle} (p_1 \geq 0 )&=& P_{{S_z(\theta)}|\chi \rangle} (p'_1 \geq 0), \nonumber\\
\label{pov-sym-rot1}
\mbox{i.e ,    } P_{{S_z(\theta)}^{\dagger}|\chi \rangle} (p_1 \geq 0 ) &=&  P_{|\chi \rangle} (p'_1 \geq 0).
\end{eqnarray}
Now, using eqn. \ref{povmform} we have from eqn. \ref{pov-sym-rot1} :
\begin{eqnarray}
 P_{|\chi \rangle} (p'_1 \geq 0 ) &=& \langle \chi|S_z(\theta)\frac{1}{2}(\alpha' I + \beta'\sigma_x) {S_z(\theta)}^{\dagger}|\chi\rangle ,\\
\mbox{i.e,  } \langle \chi| E(p'_1 \geq 0 )\chi\rangle &=&  \langle \chi|\frac{1}{2}(\alpha' I + \beta'\vec{\sigma}.\hat{n})|\chi\rangle
\end{eqnarray}
where $\hat{n} = \hat{x}cos(\theta) +\hat{y}sin(\theta) $,with  $\theta$ being the polar angle of a point  in the x-y plane. In  the last but one line we have also used 
$S_z(\theta)=exp[-i\frac{\sigma_z\theta}{2}]$.

The Hamiltonian in eqn. (\ref{model}) satisfies both the above mentioned reflection as well as the rotational symmetry properties . Choosing the detector state to be a symmetric Gaussian ,i.e, $b=a$ in eqn. (\ref{inistate2}) we 
have the required rotational invariance for all angles. Hence,  we get an approximate joint measurement of spin in any 
direction by measuring the detector momentum in that direction. 

\subsection{POVM elements}
Consider approximate joint measurement of $\sigma_x$ and $\vec{\sigma}.\hat{n}$.  We have already shown that the marginal probabilities of the joint measurement will   be given by,
\begin{eqnarray}
p(p'_1 \geq 0) &=& \langle \chi | \frac{1}{2}(1+a'\sigma_x)| \chi \rangle    \\ 
p(p'_2 \geq 0) &=& \langle \chi |\frac{1}{2}(1+a'\vec{\sigma}.\hat{n})| \chi \rangle
\end{eqnarray}
with $p'_1$ and $p'_2$ denoting momenta in $\hat{x}$ and $\hat{n}$ direction respectively ($\hat{x}$ is along positive x-axis and $\hat{n}$ along  
$ \hat{x}cos(\theta) + \hat{y}sin(\theta)$ in the momentum plane $(p_1,p_2)$). 

\subsubsection{Angular dependance of $e^0$ , $f^0$ and $g^0$}
The rotational invariance of the initial state allows one to extract the polar angular 
dependance of $e^{0}$, $f^{0}$ and $g^{0}$  (introduced in section \ref{section-ajmortho}) in the detector space ($p_1$,$p_2$) ($p_1 = pcos(\theta_p)$, $p_2=psin(\theta_p)$): 

\begin{equation}
e^0(p,\theta_p)= \frac{1}{2\pi^3a^2}\int e^{-iq_1pcos\theta_p}e^{-iq_2psin\theta_p} cos(\sqrt{q_1^ 2+q_2 ^2} )e^{-\frac{q_1^2 + q_2^2}{4a^2}} dq_1dq_2
\end{equation}
Putting, $q_1=rcos\theta$ and $q_2=rsin\theta$ we have

\begin{eqnarray}
e^0(p,\theta_p)= & C\int_0^{2\pi}\int_0^{+\infty}e^{-irpcos(\theta-\theta_p)}rcos(r) e^{\frac{-r^2}{4a^2}}drd\theta \nonumber \\
\label{enoteq1}
= & C\int_0^{+\infty}\int_0^{2\pi}e^{-irpcos(\theta-\theta_p)}d\theta g_1(r)dr
\end{eqnarray}
with C being the constant part, $g_1(r)= rcos(r) e^{\frac{-r^2}{4a^2}} $ .
Now , taking  $\theta- \theta_p= \theta'$  the theta integral in eqn. (\ref{enoteq1}) becomes,
\begin{eqnarray}
\int_0^{2\pi} e^{-irpcos(\theta-\theta_p)}d\theta &=& \int_{-\theta_p}^{2\pi-\theta_p} e^{-irpcos(\theta')}d\theta' \nonumber \\
= & \int_0^{2\pi}e^{-irpcos(\theta')}d\theta'
\end{eqnarray}
The last equation follows from the fact that the integral of a periodic function over its period is independent of the limits of integration. Thus $e^0$ is only a function of $p$: 
\begin{equation}
\label{enot}
e^0=e_1(p) 
\end{equation}
Again, similar transformation for $f^0$ yields,
\begin{equation}
\label{fnot1}
f^0(p,\theta_p)= C'\int_0^{2\pi}\int_0^{+\infty}e^{-irpcos(\theta-\theta_p)}cos(\theta)sin(r)e^{\frac{-r^2}{4a^2}}rdrd\theta
\end{equation}
with $C'$ being the constant part.
Using , $\theta- \theta_p= \theta' $  the theta integral in (\ref{fnot1}) becomes,
\begin{eqnarray*}
\int_{-\theta_p}^{2\pi-\theta_p}e^{-irpcos(\theta')}cos(\theta'+\theta_p)d\theta' = &\int_{-\theta_p}^{2\pi-\theta_p}e^{-irpcos(\theta')}cos(\theta')cos(\theta_p)d\theta' \\ 
& - \int_{-\theta_p}^{2\pi-\theta_p}e^{-irpcos(\theta')}sin(\theta')sin(\theta_p)d\theta'
\end{eqnarray*}
Now,
\begin{equation}
-\int_{-\theta_p}^{2\pi-\theta_p}e^{-irpcos(\theta')}sin(\theta')d\theta' = -\int_{0}^{2\pi}e^{-irpcos(\theta')}sin(\theta')d\theta'.
\end{equation}
Taking, $cos(\theta')=x$,  the above eqn. becomes $ \int_1^{-1} e^{-irpx}dx + \int_{-1}^{1}e^{-irpx}dx = 0  .$
Therefore, 
\begin{equation}
\label{fnot}
f^0(p,\theta_p)= f_1(p)cos(\theta_p).
\end{equation}
Proceeding exactly similarly one can show that 
\begin{equation}
\label{gnot}
g^0(p,\theta_p)= f_1(p)sin(\theta_p)
\end{equation} 

\subsubsection{Joint measurement probabilities}
\begin{figure}
\begin{center}
\label{fig-plane}
\includegraphics[height=70mm]{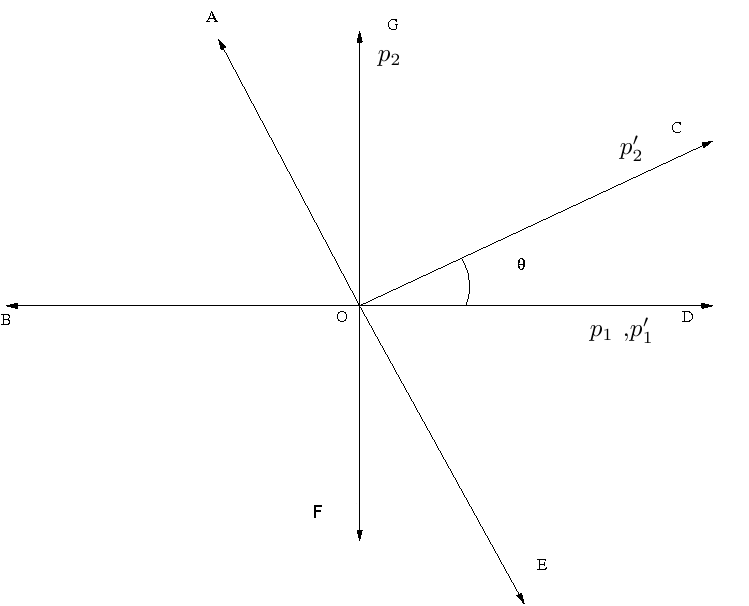}
\caption{The above figure represents the detector momentum plane. The positive $p_1$ and $p_2$ axes are represented by OD and OG respectively. OD is also the $p'_1$ axis. 
The $p'_2$ axis OC makes an angle $\theta$ with OD. By DOE , we mean here the region of the plane bounded by $OD$ and $OE$ . }
\end{center}
\end{figure}

Choosing the $p'_1$ , $p'_2$ axes according to fig. 8 , the joint probability is given by , 
\begin{equation}
\label{oblique-probability}
p(p'_1 \geq 0 , p'_2 \geq 0) =\int_{GOD}p(p_1,p_2)dp_1dp_2   +  \int_{DOE}p(p_1,p_2)dp_1dp_2  =\int_{p_1,p_2 =0}^{\infty}p(p_1,p_2)dp_1dp_2 +  \int_{DOE}p(p_1,p_2)dp_1dp_2  
\end{equation}
with, $p(p_1,p_2)$ given by (\ref{probability}) .

where the last integral represents integral on the region DOE in the figure 8. In polar coordinates the region DOE is given by, the
 set of points $\{(p,\theta'): 0 \leq p \leq \infty , -(\frac{\pi}{2}- \theta) \leq \theta' \leq 0 \}. $
So, using eqn. (\ref{probp1p2geq0}) in (\ref{oblique-probability}) we have, 
\begin{equation}
 p(p'_1 \geq 0 , p'_2 \geq 0) = \langle \chi | \left[ \frac{1}{4}I + a'\frac{\sigma_x}{4} + a'\frac{\sigma_y}{4} \right] |\chi \rangle + \int_{DOE}p(p_1,p_2)dp_1dp_2
\end{equation}

Now using the $\theta$ dependance of $e^{0}$, $f^{0}$ and $g^{0}$ in \ref{probability} from (\ref{enot}) ,  (\ref{fnot}) and (\ref{gnot}) in (\ref{probability}) we have,
\begin{eqnarray}  
\int_{DOE} p(p_1,p_2)dp_1dp_2  = &\langle \chi |( \int_{0}^{\infty} \int_{0}^{-(\frac{\pi}{2}- \theta)}[(|e_1(p)^2|+|f_1(p)^2|cos^2(\theta)+ |f_1(p)^2|sin^2(\theta))1_s \nonumber \\ 
&+ 2e_1(p)f_1(p)cos(\theta)\sigma_1 + 2e_1(p)f_1(p)sin(\theta)\sigma_2 pdpd\theta  ) |\chi \rangle
\end{eqnarray}

Now, from eqn. (\ref{probnorm}) we have,
\begin{equation}
\label{probnorm1}
\int_0^{\infty}(e_1^2 + f_1^2)2\pi pdp = 1 
\end{equation}

We also have from the definition of $a'$ (eqn. (\ref{povm-constants})),
\begin{equation}
\label{e1f1}
\int_0^{\infty} e_1f_1pdp = \frac{a'}{8} .
\end{equation}

Hence ,
\begin{equation}
p(p'_1 \geq 0 , p'_2 \geq 0) =\langle \chi |\left[ (\frac{1}{2}-\frac{\theta}{2\pi})1_s + \frac{a'}{4}(1+cos(\theta))\sigma_x + \frac{a'}{4}sin(\theta)\sigma_y \right] |\chi \rangle
\end{equation}
and similarly,
\begin{eqnarray}
p(p'_1 \leq 0 , p'_2 \geq 0) &=&\langle \chi  | \left[ \frac{\theta}{2\pi}1_s + \frac{a'}{4}(cos(\theta)-1)\sigma_x + \frac{a'}{4}sin(\theta)\sigma_y \right] |\chi \rangle \\
p(p'_1 \geq 0 , p'_2 \leq 0) &=&\langle \chi | \left[ |\frac{\theta}{2\pi}1_s + \frac{a'}{4}(-cos(\theta)+1)\sigma_x - \frac{a'}{4}sin(\theta)\sigma_y \right] |\chi \rangle \\
p(p'_1 \leq 0 , p'_2 \leq 0) &=&\langle \chi |  \left[ (\frac{1}{2}-\frac{\theta}{2\pi})1_s -\frac{a'}{4}(1+cos(\theta))\sigma_x - \frac{a'}{4}sin(\theta)\sigma_y \right] |\chi \rangle 
\end{eqnarray}

At $\theta= \frac{\pi}{2}$, we get back the POVM elements for the orthogonal case as derived in section \ref{section-ajmortho} . For $\theta \rightarrow 0$ i.e, joint measurement along two almost same 
directions, the probabilities  $p(p'_1 \leq 0 , p'_2 \geq 0)$ and $p(p'_1 \geq 0 , p'_2 \leq 0)$ vanish while $p(p'_1 \geq 0 , p'_2 \geq 0)$ and $p(p'_1 \leq 0 , p'_2 \leq 0)$ give the 
probabilities for single unsharp measurement along $\hat{x}$, as expected.

\paragraph{Remark:}
The joint measurement uncertainty relation is given by (\ref{eqajmuncer})
\begin{equation}
||\vec{a}+\vec{b}|| + ||\vec{a}-\vec{b}|| \leq 2 .
\end{equation}
If $\theta$ is the angle between $\vec{a}$ and $\vec{b}$ ,  in the case  $a=b$ (like above) one has,
\begin{equation}
\label{aleq}
a \leq \frac{1}{|sin(\frac{\theta}{2})+cos(\frac{\theta}{2})|}  .
\end{equation}
The denominator in eqn. (\ref{aleq}) is positive in $\theta \in [0,\pi]$ with maximum $\sqrt{2}$ at $\theta=\pi/2 $. This is also our bound in the approximate joint measurement we have implemented,
because for the symmetric initial detector state  case (see section VI A) approximate joint measurement happens in orthogonal directions as well. In fact we just showed that  measuring 
momenta in \textit{any} two directions yields approximate joint measurement of spin in those two directions. We saw earlier that in our case (see section VI) $a$ is able to reach about 0.628 (see fig.1a) . Thus, 
it is possible that one can have a different scheme with  same but better quality of unsharp measurement  
in both the directions. For example, for $\theta=\pi/4$ the bound by the joint measurement inequality (\ref{aleq}) is about 0.765.

\section{Spin direction fidelities}
In this section we consider the spin direction fidelities, defined in (\cite{appleby-3}) . We consider type 1 measurement as defined by the author in ref.\cite{appleby-3} in  which 
the pointer observables are taken to be the commuting components of some unit vector $\hat{n}$. The measurement scheme considered by us in section \ref{section-ajmortho} yields the momentum values $p_1$ ,$p_2$
(which can also be considered as components of $\vec{p}$). 
Alternatively, in polar coordinates we can look at the magnitude of the momentum $\vec{p}$ and its angle $\theta $ with the x-axis in the momentum plane. This angle uniquely fixes the direction of  
normalised momentum meter $\hat{p}$, and  this  is what we take for $\hat{n}$ in this section. 

The joint unsharp measurement on the system through measurement on the commuting meter observables after the meters  have interacted with the system, yields the POVM described 
by eqns. (\ref{jm-povm1}) and  (\ref{jm-povm2}) .The effect of the measurement on the system ( a completely-positive map acting on the system density matrix)  can be
described by Krauss operators for the measurement $T(\vec{p})$ . They are defined as follows 
by considering the average density matrix of the system $\rho_f$ after the measurement on the meters, considering 
all possible outcomes:

\begin{equation}
\rho_f = \int pdpd\theta \mbox{  }T(p,\theta)|\chi \rangle \langle \chi |T^{\dagger}(p,\theta) 
\end{equation}

The probability of registering the outcome at an angle $\theta$ in the ($p_1$ , $p_2$) plane is given by $p(\theta)=\langle \chi |E(\theta)|\chi \rangle $, with the angle POVM 
\begin{equation}
\label{etheta} 
E(\theta) = \int_0^{\infty} T^\dagger (p,\theta) T(p,\theta)pdp .
\end{equation}

The fidelities were verified to be  expressible in terms of E and T as, (see \cite{appleby-3})
\begin{eqnarray}
\label{etaidef}
\eta_i &=& {inf}_{|\chi \rangle \in H_{sys}} \langle \chi | \left[  \int_0^{2\pi} \frac{1}{2}(E(\theta) \vec{S}.\hat{{p}} + \vec{S}.\hat{{p}}(E(\theta))d\theta  \right] | \chi \rangle \\
\eta_f &=& {inf}_{|\chi \rangle \in H_{sys}} \langle \chi |\left[ \int T^{\dagger} \hat{n} \cdot \vec{S} T p dp d\theta \right] | \chi \rangle \\
\eta_d &=& {inf}_{|\chi \rangle \in H_{sys}} \langle \chi | \left[ \int \sum_{i=1}^3  \langle \chi \frac{1}{2}(T^{\dagger} S_i T S_i + S_iT^{\dagger} S_i T ) p dp d\theta \right]| \chi \rangle  
\end{eqnarray}

with T in our case given by,
\begin{eqnarray}
T(\vec{p})&=& \langle \vec{p} | e |\psi \rangle 1_s + \langle \vec{p} | f |\psi \rangle \sigma_x + \langle \vec{p} | g |\psi \rangle \sigma_y \nonumber\\
& = & e^{0}1_s + f^{0}\sigma_x + g^{0}\sigma_y
\end{eqnarray}

From (\ref{etheta}) we have,
\begin{eqnarray}   
E(\theta) &= & \int_0^{+\infty} (|e^0|^2 + |f^0|^2 +  |g^0|^2)pdp1_s +  \int_0^{+\infty}2Re(e^0{f^0}^*)pdp\sigma_x + \int_0^{+\infty}2Re(e^0{g^0}^*)pdp\sigma_y \nonumber\\
& = & a(\theta)1_s + b(\theta)\sigma_x + c(\theta)\sigma_y (\mbox{   say }) .
\end{eqnarray}
Hence,
\begin{equation}
\frac{1}{2}(E(\hat{n})\hat{n}.\vec{S}+ \hat{n}.\vec{S}E(\hat{n})) = \frac{1}{2}[(b(\theta)cos(\theta)+ c(\theta)sin(\theta))1_s + a(\theta)cos(\theta)\sigma_x + a(\theta)sin(\theta)\sigma_y]. 
\end{equation}

So from eqn. (\ref{etaidef}),
\begin{eqnarray}
\label{etaical}
\eta_i = &{inf}_{|\chi \rangle \in H_{sys}} \int_0^{2\pi} d\theta [\frac{1}{2}(b(\theta)cos(\theta)+ c(\theta)sin(\theta)) + \frac{1}{2}a(\theta)cos(\theta)\langle \chi|\sigma_x|\chi \rangle  + \frac{1}{2}a(\theta)sin(\theta)\langle \chi|\sigma_y |\chi \rangle] \nonumber\\
 = & \int_0^{\infty}\int_0^{2\pi}e^0{f^0}^*cos(\theta)+e^0{g^0}^*sin(\theta)pdpd\theta + \int_0^{\infty}\int_0^{2\pi} (|e^0|^2 + |f^0|^2 +  |g^0|^2)cos(\theta)pdpd\theta \langle \chi|\sigma_x|\chi \rangle \nonumber\\ 
   & +\int_0^{\infty}\int_0^{2\pi}(|e^0|^2 + |f^0|^2 +  |g^0|^2)sin(\theta)pdpd\theta  \langle \chi|\sigma_y|\chi \rangle 
\end{eqnarray}
Changing back to Cartesian coordinates we have,
\begin{equation}
\label{etai}
\eta_i =  \int_{-\infty}^{\infty} e^0{f^0}^* \frac{p_1}{(p_1 ^ 2 + p_2 ^2 ) ^ \frac{1}{2}}dp_1dp_2 + \int_{-\infty}^{\infty} e^0{g^0}^* \frac{p_2}{(p_1 ^ 2 + p_2 ^2 ) ^ \frac{1}{2}}dp_1dp_2 .
\end{equation}
The last two terms on the RHS of eqn. (\ref{etaical})  vanish because the 3rd term is  is odd in $p_1$ and the 4th one is odd in $p_2$ .

Now using (\ref{e1f1}) we have,
\begin{eqnarray}
\eta_i &= &\int e^0f^0cos(\theta)pdpd\theta + \int e^0g^0sin(\theta)pdpd\theta \nonumber\\
& = & \int e_1f_1pdp \int_{\theta=0}^{2\pi} cos^2(\theta)d\theta + \int e_1f_1pdp \int_{\theta=0}^{2\pi} sin^2(\theta)d\theta \nonumber\\
& = &2\pi\int e_1f_1pdp 
\end{eqnarray}
Thus,
\begin{equation}
\label{etai-a'}
\eta_i= \frac{\pi}{4}a'
\end{equation}

\paragraph{\textbf{Bound on $a'$:}}
It was shown in \cite{appleby-3} that  $\eta_i$ is bounded by the value $s=\frac{1}{2}$ . From here we get the bound on $a'$ to be about .64 which is almost exactly what is got in the simulation of $a'$ (0.6292 in fig. 1a). 
 The joint measurement uncertainty relation (\ref{ajm-ortho}) allows $a'$ to go till .707. This is because eqn. (\ref{ajm-ortho}) refers to the most general 
approximate joint measurement in orthogonal directions without reference to any Arthur-Kelly kind of an implementation.  Also eqn. (\ref{etai-a'}) shows that the error of 
retrodiction (given in eqn. (\ref{retroerror})) falls as the measurement becomes  
sharp.

A similar calculation shows that,
\begin{eqnarray}
\eta_f = &{inf}_{|\chi \rangle \in H_{sys}}\int p d\theta dp \langle\chi |[[cos(\theta) (|e^0|^2 + |f^0|^2 -  |g^0|^2)+ 2Re(f^0{g^0}^*)sin(\theta)]\sigma_x + [sin(\theta) (|e^0|^2 - |f^0|^2 \nonumber\\ 
 & +|g^0|^2) + 2Re(f^0{g^0}^*)cos(\theta)]\sigma_y + [e^0f^0cos(\theta) + e^0g^0sin(\theta)]1_s] |\chi \rangle
\end{eqnarray} 
and,
\begin{equation}
\eta_d = {inf}_{|\chi \rangle \in H_{sys}}\int p d\theta dp \langle\chi \left[ |\frac{3}{4}(|e^0|^2 + |f^0|^2 +  |g^0|^2)1_s + \frac{e^0{f^0}^*}{2}\sigma_x + \frac{e^0{g^0}^*}{2}\sigma_y \right]| \chi\rangle .
\end{equation}

Thus, as in the case of original Arthur-Kelly model, the fidelities are independent of the initial system state\cite{HBK-1}.

Again, because of the parity of various terms present there $\eta_f$ turns out to be the same as $\eta_i$ . 

From the probability normalisation condition given in eqn. (\ref{probnorm}) we have $\eta_d = \frac{3}{4}$, \textit{independent of initial apparatus state and $\eta_i$ ,$\eta_f$ . } 

This shows that any error-disturbance relationship
between error of retrodiction/ prediction and error of disturbance does not hold for all choices of the direction pointer observable. 

\section{Approximate joint measurement for three qubit observables}
An approximate joint measurement observable for three unsharp qubit observables is defined by extending conditions given in (\ref{jointpovm}) in the obvious way. Unlike the 
two  observable case no necessary-sufficient condition 
for the approximate joint measurement on the marginal effect parameters in  $ {\cal{R}}^4$  (like the ones derived for example in \cite{oh}) is known. No measurement uncertainty relation like equation (\ref{mur})  
, but which is stronger than the relations applied pairwise is known. Here,  we derive a necessary condition on the approximate joint measurement of three qubit observables which yields the familiar 
necessary-sufficient condition known in literature for the case of mutually orthogonal unbiased observables (to be discussed below). We show that it holds 
even when the observables are biased.  Also when one of the measurements 
is trivial  i.e the corresponding marginal effect is $\frac{I}{2}$ representing equiprobable guesses of the values of the observable , our condition reproduces eqn. 
(\ref{eqajmuncer}). 

The eight joint effects $G_{+++}$, $G_{++-}$,..........,$G_{---}$  have to satisfy six marginality conditions like  
\begin{equation}
\Upsilon^1_+ = G_{+++} + G_{++-} + G_{+-+} + G_{+--} 
\end{equation}
A joint measurement of three observables gives rise to three two-observable joint measurements. Let $G^{12}_{++}$ denote the joint measurement marginal effect corresponding to the outcome (++) in directions 1 and 2. Then seven of the  the effects
($G_{+++}$,$G_{++-}$ , etc. )  can be written in terms of the three single-observable marginal effects , three pairwise joint measurement marginal effects  and $G_{+++}$, as follows :
\begin{eqnarray}
G_{++-}  &=&  G^{12}_{++} - G_{+++}, \\
G_{-+-}  &=&  G^{23}_{+-} - G^{12}_{++} + G_{+++} ,\\
G_{--+}  &=&  G^{13}_{-+} + G^{23}_{+-} + G_{+++} - \Upsilon^2_+ ,\\
G_{-++}  &=&  \Upsilon^2_+ - G^{23}_{+-} - G_{+++} ,\\
G_{+-+}  &=&  G^{13}_{++} - G_{+++} ,\\
G_{---}  &=&  G^{23}_{++} + G^{12}_{--} - G^{13}_{-+} - G_{+++} ,\\
G_{+--}  &=&  G^{12}_{+-} - \Upsilon^3_+ + G^{13}_{-+} + G_{+++} .
\end{eqnarray}
It was shown in \cite{oh} that any set of effects for joint measurement of two observables (defined in eqn.(\ref{jointpovm})) can be written in the form

\begin{equation}
\label{gfortwo}
G_{sgn(a)sgn(b)}(Z,\vec{z}) = \frac{(1+ ax + by+ abZ)I + (ab\vec{z} + a\vec{m} + b\vec{n}).\vec{\sigma}}{4}  
\end{equation} 
with $a,b \in \{1,-1\}$, all other scalars in $\cal{R}$, all vectors in ${\cal{R}}^3 $ and the two single-qubit marginal effects given by ,
$\hat{\Upsilon}_{\pm}(x,\vec{m}) = \frac{1}{2}((1 \pm x)I \pm \vec{m}.\vec{\sigma}) $ and  
$\hat{\Upsilon}_{\pm}(y,\vec{n}) = \frac{1}{2}((1 \pm y)I \pm \vec{n}.\vec{\sigma}) $ 
 with positivity constraint 
being $|x|+m \leq 1$ and $|y|+n \leq 1$ . Condition \ref{gfortwo}  is true because for the two observable case all the effects can be expressed in terms of one effect (say $G_{++}$) and 
the marginals (\cite{ajm}) . The freedom in $(Z,\vec{z})$ suffices to specify an arbitrary $G_{++}$. 

      It then follows from equation \ref{gfortwo}  that  the set of joint effects for the three observable case is of the form,
\begin{eqnarray}           
\label{gforthree}
G_{sgn(a)sgn(b)sgn(c)}([Z], [\vec{z}]) =& ((1+ax+by+cz+abZ_1 + bcZ_2 + caZ_3 + abcZ_4)I + (ab\vec{z_1} + bc\vec{z_2} + ca\vec{z_3}  \nonumber \\
& + abc\vec{z_4}+a\vec{l} + b\vec{m}+ c\vec{n}).\vec{\sigma})/8
\end{eqnarray}
with, $a,b,c \in \{1,-1\}$, the two-marginals $G_{12}= G(\vec{z_1}, Z_1)$ , $G_{23}= G(\vec{z_2}, Z_2)$ , $G_{13}= G(\vec{z_3}, Z_3)$ given by 
eqn. (\ref{gfortwo}), and the one marginals being
 $\hat{\Upsilon}_{\pm}(x,\vec{l}) = \frac{1}{2}((1 \pm x)I \pm \vec{l}.\vec{\sigma})$ , 
$\hat{\Upsilon}_{\pm}(y,\vec{m}) = \frac{1}{2}((1 \pm y)I \pm \vec{m}.\vec{\sigma}) $ , 
$\hat{\Upsilon}_{\pm}(z,\vec{n}) = \frac{1}{2}((1 \pm z)I \pm \vec{n}.\vec{\sigma}) $. Given the positivity of the one marginals 
, the positivity of the eight joint effects
 places restrictions on the marginal effect parameters  which are interpreted as measurement uncertainty relations  in contrast to the usual state uncertainty relations.

\subsection{Necessary condition}
From eqn. (\ref{gforthree}) ,
$G_{+++} \geq 0$ gives,
\begin{equation}
\label{eqi}
||\vec{l} + \vec{m} + \vec{n}+ \vec{z_1} + \vec{z_2}+ \vec{z_3} + \vec{z_4}|| \leq 1+x+y+z+Z_1 + Z_2 + Z_3 + Z_4 
\end{equation}  
 while $G_{---} \geq 0$ gives,
\begin{equation}
\label{eqii}
||-\vec{l} - \vec{m} - \vec{n}+ \vec{z_1} + \vec{z_2}+ \vec{z_3} -\vec{z_4}|| \leq 1-x-y-z+Z_1 + Z_2 + Z_3 - Z_4 
\end{equation}
Equations (\ref{eqi}) and (\ref{eqii}) together can be interpreted as  the collection of all points $(\vec{z_1} + \vec{z_2} + \vec{z_3})\in  {\cal{R}}^3 $, each of which
lies within two spheres in  ${\cal{R}}^3$   with their centers at  $(\vec{l} + \vec{m} + \vec{n}+ \vec{z_4})$ and $-(\vec{l} + \vec{m} + \vec{n}+ \vec{z_4})$ and radii 
    $ (1-x-y-z+Z_1 + Z_2 + Z_3 - Z_4)$, $(1+x+y+z+Z_1 + Z_2 + Z_3 + Z_4)$ respectively. Thus, the distance between their centers should be less than or equal to the sum of their radii implying, 
\begin{equation}
\label{eqiii}
||-\vec{l} - \vec{m} - \vec{n}- \vec{z_4}|| \leq 1 + Z_1 + Z_2 + Z_3
\end{equation}

Similarly, by consideration of the other complementary pairs $G_{pqr}$ and $G_{(-p)(-q)(-r)}$ for $p,q,r = \pm$ we get three other equations,
\begin{eqnarray}
\label{inequality1}
||\vec{l} + \vec{m} - \vec{n}- \vec{z_4}|| &\leq &1 + Z_1 - Z_2 - Z_3, \\
||\vec{l} - \vec{m} + \vec{n}- \vec{z_4}|| &\leq & 1 - Z_1 - Z_2 + Z_3, \\
\label{eqiv}
\label{inequality4}
||-\vec{l} + \vec{m} + \vec{n}- \vec{z_4}||&\leq & 1 - Z_1 + Z_2 -Z_3 . 
\end{eqnarray}

\subsection{Geometric interpretation}
\begin{figure}
\begin{center}
\includegraphics[height=70mm]{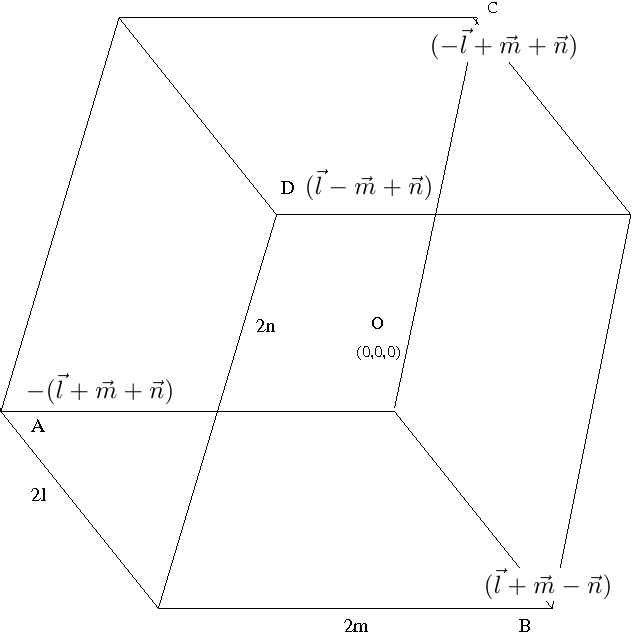}
\end{center}
\caption{Geometric interpretation of inequalities (\ref{inequality1}) -  (\ref{inequality4}) . Solid spheres about the centers A (with $\vec{OA}= -(\vec{l} + \vec{m} + \vec{n})$) , B (with $\vec{OB}= (\vec{l} + \vec{m} - \vec{n})$),
 C (with $\vec{OC} = (-\vec{l} + \vec{m} + \vec{n})$) and D (with $\vec{OD}= (\vec{l} - \vec{m} + \vec{n})$) and radii $(1 + Z_1 + Z_2 + Z_3)$ ,  $(1 + Z_1 - Z_2 - Z_3)$ , $ (1 - Z_1 - Z_2 + Z_3)$, $ (1 - Z_1 + Z_2 -Z_3)$   
respectively have to intersect so that the sum of their radii is 4.}            
\label{fig-cuboid}
\end{figure}

The above inequalities (\ref{eqiii}) -  (\ref{inequality4})  imply that we have a cuboid with edges $2\vec{l}$ , $2\vec{m}$ and $2\vec{n}$  with  origin at the centre of this cuboid and solid spheres 
about points A ($-\vec{l} - \vec{m} - \vec{n}$) ,  B($\vec{l} + \vec{m} - \vec{n}$), C ($-\vec{l} + \vec{m} + \vec{n} $) and  D  ($\vec{l} - \vec{m} + \vec{n} $)have to intersect
 so that the sum of the radii of the spheres is 4 (see fig. 5). That is we necessarily have:

\begin{equation}
\label{eqiv}
||-\vec{l} - \vec{m} - \vec{n}- \vec{z_4}||+||\vec{l} + \vec{m} - \vec{n}- \vec{z_4}||+||\vec{l} - \vec{m} + \vec{n}- \vec{z_4}||+||-\vec{l} + \vec{m} + \vec{n}- \vec{z_4}|| \leq 4 
\end{equation}

Now, for any given  set of points in $ {\cal{R}}^3 $, the Fermat-Toricelli (F-T) point of the set is defined to be the point which minimises the sum of distances from that point in $ {\cal{R}}^3 $ 
to the points of the set.
The F-T point  has been studied for quite long and its properties for any set of four non-coplanar points are well known. (\cite{fermat}). 
For example, it is known that the F-T point either belongs to the set itself or else it is the point at which the gradient of the sum of distances vanish. 
Thus choosing $\vec{z_4}$ to be the F-T point of the set of points A,B,C,D one necessarily has  (according to the definition of the F-T point),
\begin{equation}
\label{fermat-condition}
||-\vec{l} - \vec{m} - \vec{n}- {\vec{z_4}}_F||+||\vec{l} + \vec{m} - \vec{n}- {\vec{z_4}}_F||+||\vec{l} - \vec{m} + \vec{n}- {\vec{z_4}}_F||+||-\vec{l} + \vec{m} + \vec{n}- {\vec{z_4}}_F|| \leq 4 , 
\end{equation}
where  $\vec{z_4}_F$ denotes the F-T  point.

\subsection{Sufficiency }
In order to evaluate the F-T point for a particular set of points {A,B,C,D} defined above we will use the following theorem from ref. \cite{fermat} (Theorem 2.1),  due to Lorentz Lindelf and Sturm. 

\textbf{Theorem 2:} Suppose  $\vec{z_4}$ is the F-T point for a set of points $S_n \equiv \{ \vec{a_i} \in {\cal{R}}^3 : i=1,2,...n \}.$ Then  either  $\vec{z_4}$ belongs to the set $S_n$ or $\vec{z_4} \notin S_n$ .      

i) If $\vec{z_4} \in S_n$ then for $\vec{z_4} = \vec{a_j}$ , for some $j \in \{1,2,...n\},$ with  $|| \sum_{i=1(i\neq j)}^n \frac{(\vec{a_i}- \vec{a_j})}{||\vec{a_i}- \vec{a_j}||} || \leq 1 .$ 

ii) If $\vec{z_4} \notin S_n$ then $\vec{z_4}$ is the point at which  $||\sum_{i}^{n} \frac{(\vec{a_i}- \vec{z_4})}{||\vec{a_i}- \vec{z_4}||} || = 0 $

Thus, for the first case the resultant of unit  vectors to the FT  point  from other points of the set has magnitude less than 1.  In the second case the unit vectors  
from the F-T point to the points of the set add up to the null vector. The condition for the second case is also the condition for the gradient of the function representing the sum of distances
from $\vec{z_4}$ to $\vec{a_i}$, for $i=1,2,...n$, to vanish at $\vec{z_4}$.

\subsection*{$\vec{l}$, $\vec{m}$, $\vec{n}$ mutually orthogonal}
Suppose we are considering the case of approximate joint measurement in three orthogonal directions $\vec{l}$ , $\vec{m}$ and $\vec{n}$. In this case we have,
$||-\vec{l} - \vec{m} - \vec{n}||=||\vec{l} + \vec{m} - \vec{n}||= ||\vec{l} - \vec{m} + \vec{n}||=||-\vec{l} + \vec{m} + \vec{n}||$. Hence, at the origin($\vec{z_4}=\vec{0}$) 
of the cuboid (see fig. (\ref{fig-cuboid})) we have,  
\begin{align}
\sum_{i}\frac{(\vec{a_i}- \vec{z_4})}{||\vec{a_i}- \vec{z_4}||} = (-\vec{l} - \vec{m} - \vec{n} +\vec{l} - \vec{m} + \vec{n}+ \vec{l} + \vec{m} - \vec{n}-\vec{l} + \vec{m} + \vec{n})/(l^2+m^2+n^2)=0 
\end{align}
Hence equation  (\ref{fermat-condition})  gives,
\begin{equation}
\label{nec-suff}
l^2 + m^2 + n^2 \leq 1
\end{equation}
This condition was shown to be sufficient in \cite{busch-unsharp} . The necessity of this condition was shown in \cite{necessity-erica} assuming unbiasedness 
of the marginals (i.e  $x=y=z=0$ in eqn. (\ref{gforthree})) by considering measurements by two parties on a singlet state. We have shown above that this is true for biased 
cases as well. Sufficiency of the above condition  (\ref{nec-suff}) shows that condition  (\ref{fermat-condition}) is sufficient for the case of approximate joint  measurement in 
three orthogonal directions. Also note that condition (\ref{nec-suff}) is stronger than  pairwise conditions for two-observable joint measurement in orthogonal directions like (\ref{ajm-ortho}) 
which when added produces the bound of 1.5 on the lhs of  equation (\ref{nec-suff}). 

\subsection*{Reduction to two-observable inequality} 
Suppose our joint measurement scheme is such that some approximate joint measurement on two observables is performed while the value of the third observable is guessed
 with probability of + being $\frac{1 + z }{2}$ and that of - being $\frac{1 - z }{2}$. This will correspond to $\vec{n}=0$ i.e the corresponding marginal $\Upsilon_{\pm}(z,\vec{n}) = \frac{(1 \pm z)I }{2} $ .
In this case the points A($-\vec{l} - \vec{m}$) , B($\vec{l} + \vec{m}$ ), C($\vec{l} - \vec{m} $)  and D ($-\vec{l} + \vec{m} $) form a parallelogram of length $|2\vec{l}|$ and $|2\vec{m}|$ about the
origin O . OA and OB lie opposite to each other and so does OC and OD. Thus, condition (ii) of the theorem 2 is satisfied and the 
origin is the F-T point. Hence, condition (\ref{fermat-condition})  reproduces equation (\ref{eqajmuncer}) which shows that the bounds on the unsharpnesses  is not more 
stringent than the two -observable case, as is to be expected.

\subsection{Arthur-Kelly model}
The Arthur-Kelly model for the case for three qubits proceeds exactly as that of two qubits with the Hamiltonian of the measurement interaction of the form,
$H= -( q_1 \otimes \sigma_x + q_2 \otimes \sigma_y+ q_3 \otimes \sigma_z)$. The nature of the interaction is impulsive as in eqn. (\ref{model}) and the corresponding unitary evolution is 
given by,
\begin{equation}
U= e(q_1,q_2,q_3)\otimes \sigma_x +f(q_1,q_2,q_3)\otimes \sigma_y +g(q_1,q_2,q_3)\otimes \sigma_z   .
\end{equation}
with,
\begin{eqnarray}   
 e(q_1,q_2,q_3) & = & cos (\sqrt{q_1 ^ 2 + q_2 ^2+ q_3^2 }) \\
f(q_1 , q_2, q_3) & = & iq_1 \frac{sin (\sqrt{q_1 ^ 2 + q_2 ^2+ q_3^2 })}{\sqrt{(q_1 ^ 2 + q_2 ^2+q_3^2 )}}, \\
g(q_1 , q_2,q_3) & = & iq_2  \frac{sin (\sqrt{q_1 ^ 2 + q_2 ^2+ q_3^2 })}{\sqrt{(q_1 ^ 2 + q_2 ^2+ q_3^2 )}}, \\
h(q_1 , q_2,q_3) & = & iq_3  \frac{sin(\sqrt{q_1 ^ 2 + q_2 ^2+ q_3^2 })}{\sqrt{(q_1 ^ 2 + q_2 ^2+ q_3^2 )}}. 
\end{eqnarray}
The initial joint state of the three detectors and system is  $ |{\psi}_{in} \rangle = |{\psi}_1 \rangle \otimes |{\psi}_2 \rangle \otimes  |{\psi}_3 \rangle \otimes |{\chi}\rangle $. As before, the initial detector states are
 Gaussians: $\psi_1(q_1)= (\frac{1}{\sqrt{2\pi}}e^-{\frac{q_1^2}{2a^2}})^\frac{1}{2}$ ,  $\psi_2(q_2)= (\frac{1}{\sqrt{2\pi}}e^-{\frac{q_2^2}{2b^2}})^\frac{1}{2}$ and
 $\psi_3(q_3)= (\frac{1}{\sqrt{2\pi}}e^-{\frac{q_3^2}{2c^2}})^\frac{1}{2}$.

 The detector outcome ($ (p_1 \geq 0 , p_2 \geq 0, p_3 \geq 0$ ) is taken here to correspond to the outcome  (+,+,+) for the joint unsharp measurement 
of the system observables  $\sigma_x$, $\sigma_y$, $\sigma_z$ ; ($p_1 \geq 0 , p_2 \geq 0, p_3 \leq 0 $) to (+,+,-)
and so on.
The POVM elements corresponding to the outcomes (+++), (++-) , etc. of the joint unsharp measurement of $\sigma_x$ , $\sigma_y$ and $\sigma_z$ have a similar structure to 
that for the case of two observables :  $(+++) \leftrightarrow \frac{1}{8}(I+\frac{a'}{8}\sigma_x + \frac{b'}{8}\sigma_y + \frac{c'}{8}\sigma_z)$ ,
$(++-) \leftrightarrow \frac{1}{8}(I+\frac{a'}{8}\sigma_x + \frac{b'}{8}\sigma_y - \frac{c'}{8}\sigma_z)$ ,...... $(---) \leftrightarrow \frac{1}{8}(I-\frac{a'}{8}\sigma_x - \frac{b'}{8}\sigma_y - \frac{c'}{8}\sigma_z)$ with,
\begin{align}
\label{a'}
a' & = \int_{p_1=0}^{+\infty} \int_{p_2=-\infty}^{+\infty} \int_{p_3=-\infty}^{+\infty}4(f^0e^0) dp_1dp_2dp_3 , \\
b' & = \int_{p_2=0}^{+\infty} \int_{p_1=-\infty}^{+\infty}\int_{p_3=-\infty}^{+\infty} 4(g^0e^0) dp_1dp_2dp_3 ,\\                       
c' & = \int_{p_3=0}^{+\infty} \int_{p_1=-\infty}^{+\infty}\int_{p_2=-\infty}^{+\infty} 4(h^0e^0) dp_1dp_2dp_3 .
\end{align}
with $e^0 , f^0, g^0, h^0$ being the fourier transforms of $e,f, g, h$ .
\subsubsection{Symmetric case with a=b=c}
With the initial joint  pointer state  a symmetric Gaussian, as in the two observable case, we have for the unsharp measurement along any three directions  $\hat{n} ,\hat{m}, \hat{l} $, the marginal effects given by
$\Upsilon^{\hat{n}}_{+} = \frac{1}{2}(I+a'\hat{n}.\vec{\sigma})$ , $\Upsilon^{\hat{m}}_{+} = \frac{1}{2}(I+a'\hat{m}.\vec{\sigma})$ and $\Upsilon^{\hat{l}}_{+} = \frac{1}{2}(I+a'\hat{l}.\vec{\sigma})$ . 
Thus from eqn. (\ref{nec-suff}) we have $a'$ to be bounded by $\frac{1}{\sqrt(3)}=0.577$ . Numerically , for our scheme, $a'$  is seen to be able to reach up to about 0.49 . It cannot reach the
bound 0.577,as for two-observable joint measurement.(see fig. 6).

For approximate joint measurement in the directions $\hat{l} = (1,0,0)$ , $\hat{m} = (cos(\phi) , sin(\phi) , 0)$ , $\hat{n} = (sin(\theta) cos(\phi_1) , sin(\theta) sin(\phi_1) , cos(\theta))$ with $\theta=0.414 \pi$,   
$\phi_1=0.159 \pi$ and $\phi=0.477 \pi$, the Fermat- Toricelli point of the set of points A,B,C,D, as defined before, is seen to be at $\vec{l} + \vec{m} - \vec{n}$ . This yields, from inequality (\ref{fermat-condition}), 
$a' \leq 0.667$.
Thus, as for the two observable case,  considerably more freedom is available for unbiased measurement in non-orthogonal directions which our scheme cannot take advantage of.  

\begin{figure}
\label{aprimesym}
\includegraphics[height=70mm]{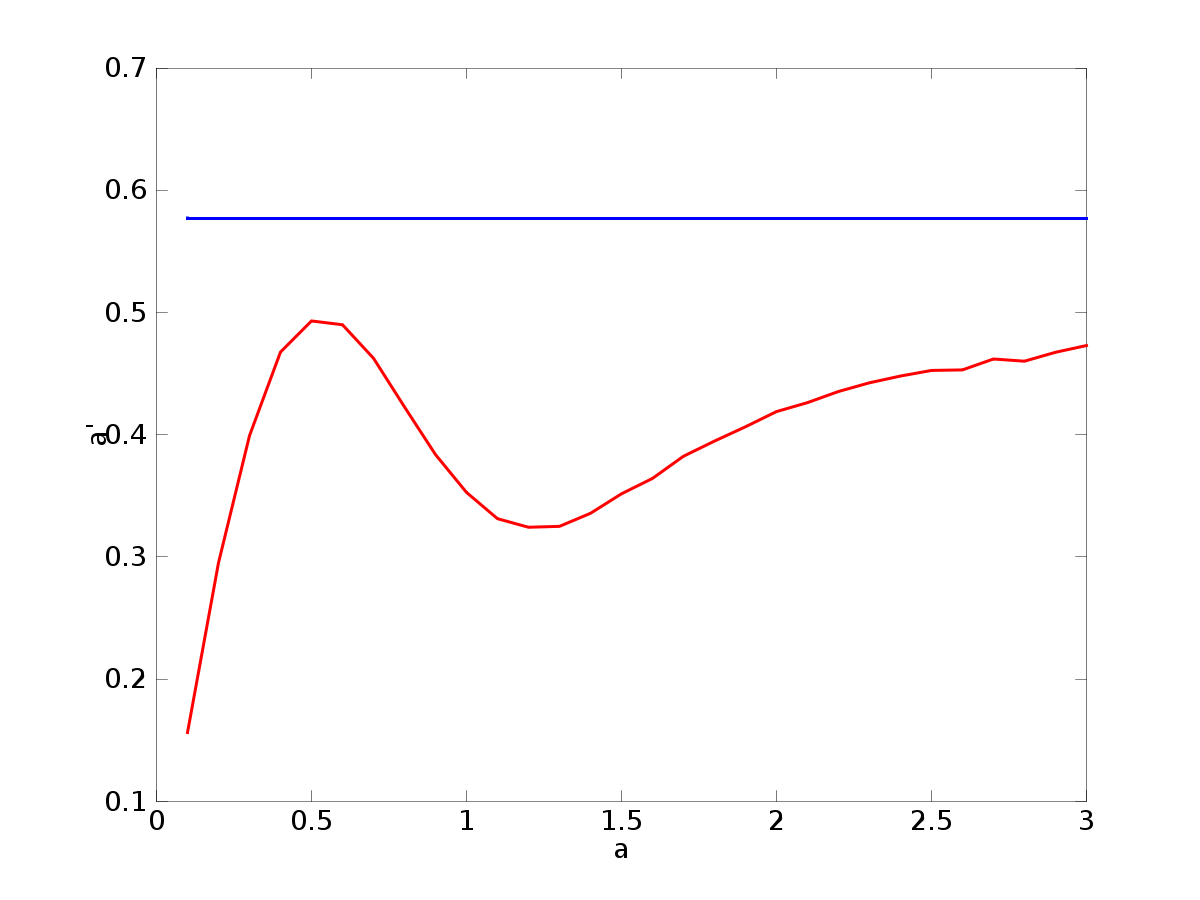}
\caption{ A plot of $a'$ vrs $a$ for symmetric Gaussian initial state }
\end{figure}

\section{Conclusions}
Non-commuting observables cannot be measured  jointly. However it is possible in the POVM formalism to do joint measurements of (generally non-commuting) observables  which are approximations of the actual non-commuting 
observables and called unsharp observables. 
We have considered in this paper approximate joint measurement of two and three qubit observables separately, through an Arthur-Kelly like model for  qubit observables.This model comes naturally when one considers a 
Stern-Gerlach setup with a linear magnetic field. In the Stern-Gerlach setup the momenta of the atoms  acts as pointer observables for their spin degrees of freedom. 
Considering  approximate joint measurement of $\sigma_x$ and $\sigma_y$ 
through this model, we have shown here numerically that the measurement uncertainty relations derived elsewhere(see ref.\cite{ajm}) hold. It has also been shown how increasing the relative sharpness  of the initial momentum wavefunctions of the 
detectors leads measurement of one observable to become almost sharp while that of the other to become almost trivial. 
Effect of initial detector states on the post-measurement state of the system has also been considered. The action of the measurement interaction on the system
turns out to be that of an asymmetric depolarising channel. This forms the basis of a physical understanding of the origin of complementarity (between $\sigma_x$
and $\sigma_y$ measurement )in the model. We also see an indication
of the entanglement between the system and detectors increasing as one of the  measurements becomes sharper. 

      We have considered two different characterisations of unsharpness. Firstly, by comparing the probability distribution of the values of observable to be approximately measured with that of the approximate observable. Secondly, by considering 
the alignment of the momentum direction with the spin observable in the Heisenberg picture. We have shown that for the case in which both the observables are approximated equally well, the corresponding measures of unsharpness 
are proportional. For our choice of pointer observable the measures checking the alignment and disturbance due to measurement do not seem to satisfy an error-disturbance relationship.

     We have expounded the connection between the symmetries of the underlying Hamiltonian for measurement interaction and initial detector states with the  joint measurement in a lemma.This was first stated by Martens et al.(\cite{Myunck}). It has 
then been used  to perform approximate joint measurement in arbitrary directions. The POVM elements  calculated for the same match with that which were found earlier in the orthogonal case. They also turn as expected  to that of a single
unsharp measurement when the directions are taken to be almost same.   
    
       For the case of joint unsharp measurement of three qubit observables we have given a necessary condition to be satisfied by the  parameters of the marginal POVM elements . This condition  has been derived from certain geometric
considerations involving the so called Fermat-Toricelli point . The condition is sufficient for the case of three orthogonal observables and  identical to the only necessary-sufficient condition known for three 
-observable joint measurements.  Our proof shows that this also holds for biased unsharp measurements , namely those in which the probability of obtaining 'up' or 'down' is different for a maximally mixed state. 

      The measurement scheme employed by  us for joint measurement in non-orthogonal directions cannot take advantage of the greater freedom available for better approximation  compared to the orthogonal case. It will
be interesting to see such a scheme in the Arthur-Kelly setup that is able to get close to the bound set by eqn.(\ref{eqajmuncer}) for arbitrary directions. We have shown that the retrodictive fidelity $\eta_i$ and the unsharpness $a'$ 
are proportional for the symmetric joint unsharp measurement case that we have considered. It may be possible to connect the two pictures in a more general setting starting with certain  symmetries of the Hamiltonian and initial 
detector states. The problem of determining  necessary-sufficient conditions for the most general joint measurement of three observables by extension of our approach or otherwise is also left open. 
 
\section*{Acknowledgments}
We want to thank  Syed Raghib Hasaan and Rajeev Singh for useful discussions.

\end{document}